\documentclass[twocolumn,twocolappendix]{aastex631}
\usepackage{hyperref,graphicx}

\begin{document}

\title{Spectroscopic Imaging of the Sun with MeerKAT: Opening a New Frontier in Solar Physics}

\author[0000-0001-8801-9635]{Devojyoti Kansabanik}
\affiliation{National Centre for Radio Astrophysics, Tata Institute of Fundamental Research, S. P. Pune University Campus, Pune 411007, India}

\author[0000-0002-2325-5298]{Surajit Mondal}
\affiliation{Center for Solar-Terrestrial Research, New Jersey Institute of Technology, 323 M L King Jr Boulevard, Newark, NJ 07102-1982, USA}

\author[0000-0002-4768-9058]{Divya Oberoi}
\affiliation{National Centre for Radio Astrophysics, Tata Institute of Fundamental Research, S. P. Pune University Campus, Pune 411007, India}

\author[0000-0002-9875-7436]{James O. Chibueze}
\affiliation{Department of Mathematical Sciences, University of South Africa, Cnr Christian de Wet Rd and Pioneer Avenue, Florida Park, 1709, Roodepoort, South Africa}
\affiliation{Centre for Space Research, Physics Department, North-West University, Potchefstroom 2520, South Africa}
\affiliation{Department of Physics and Astronomy, Faculty of Physical Sciences, University of Nigeria, Carver Building, 1 University Road, Nsukka 410001, Nigeria}

\author[0000-0003-3659-7956]{N. E. Engelbrecht}
\affiliation{Centre for Space Research, Physics Department, North-West University, Potchefstroom 2520, South Africa}

\author[0000-0002-0205-0808]{R. D. Strauss}
\affiliation{Centre for Space Research, Physics Department, North-West University, Potchefstroom 2520, South Africa}

\author[0000-0002-8078-0902]{E. P. Kontar}
\affiliation{School of Physics and Astronomy, University of Glasgow, Glasgow, G12 8QQ, UK}

\author[0000-0002-5915-697X]{G. J. J. Botha}
\affiliation{Department of Mathematics, Physics, and Electrical Engineering, Northumbria University, Newcastle upon Tyne, NE1
8ST, UK}

\author[0000-0003-2099-8093]{P. J. Steyn}
\affiliation{Centre for Space Research, Physics Department, North-West University, Potchefstroom 2520, South Africa}, 

\author[0000-0001-6917-1105]{Amor\'{e} E. Nel}
\affiliation{The South African National Space Agency, Hermanus, 7200, South Africa}

\accepted {November 4, 2023}
\published {January 17, 2024}

\begin{abstract}
Solar radio emissions provide several unique diagnostics to estimate different physical parameters of the solar corona, which are otherwise simply inaccessible. However, imaging the highly dynamic solar coronal emissions spanning a large range of angular scales at radio wavelengths is extremely challenging. At GHz frequencies, MeerKAT radio telescope is possibly globally the best-suited instrument at present for providing high-fidelity spectroscopic snapshot solar images. Here, we present the first published spectroscopic images of the Sun made using the observations with MeerKAT in the 880 -- 1670 MHz band. This work demonstrates the high fidelity of spectroscopic snapshot MeerKAT solar images through a comparison with simulated radio images at MeerKAT frequencies. The observed images show extremely good morphological similarities with the simulated images. Our analysis shows that below $\sim$900 MHz MeerKAT images can recover essentially the entire flux density from the large angular scale solar disc. Not surprisingly, at higher frequencies, the missing flux density can be as large as $\sim$50\%. However, it can potentially be estimated and corrected for. We believe once solar observation with MeerKAT is commissioned, it will enable a host of novel studies, open the door to a large unexplored phase space with significant discovery potential, and also pave the way for solar science with the upcoming Square Kilometre Array-Mid telescope, for which MeerKAT is a precursor.
\end{abstract}


\section{Introduction}\label{sec:intro}
Since the discovery of solar radio emission \citep{reber1944}, the Sun has been studied in great detail in a wide range of frequencies spanning the range from a few kHz to several hundreds of GHz \citep[e.g.][]{Pick2008,Gary2023_review}. Despite this long history of observations and studies, the Sun still harbors several mysteries. Improved observations from the new telescopes enabled by technological advances help solve these mysteries. At the same time, these new advancements probe the Sun in ways not possible earlier and, thus far, have invariably opened up a very rich discovery space. Interesting results coming from new instruments like the Solar Orbiter \citep{muller2020,Garc2021}, Parker Solar Probe \citep{Raouafi2023}, Daniel K. Inouye Solar Telescope \citep[DKIST,][]{dkist2020,DKIST_2021}, Murchison Widefield Array \citep[MWA,][]{lonsdale2009,tingay2013,Wayth2018}, LOw Frequency ARray \citep[LOFAR,][]{lofar2013}, Expanded Owens Valley Solar Array \citep[EOVSA,][]{Gary2012}, the NenuFAR \citep{Zarka2018,Briand2022}, Long Wavelength Array \citep[LWA;][]{kassim2010}, the Owens Valley Long Wavelength Array \citep[OVRO-LWA;][]{Hallinan2023} are testament to this. Except for EOVSA, these new-generation radio telescopes are not dedicated solar facilities, nonetheless, they have already been opening up large expanses of pristine unexplored phase space and making substantial contributions. 

\begin{figure*}[!htbp]
\centering
\includegraphics[trim={2.3cm 0cm 2cm 0cm},clip,scale=0.75]{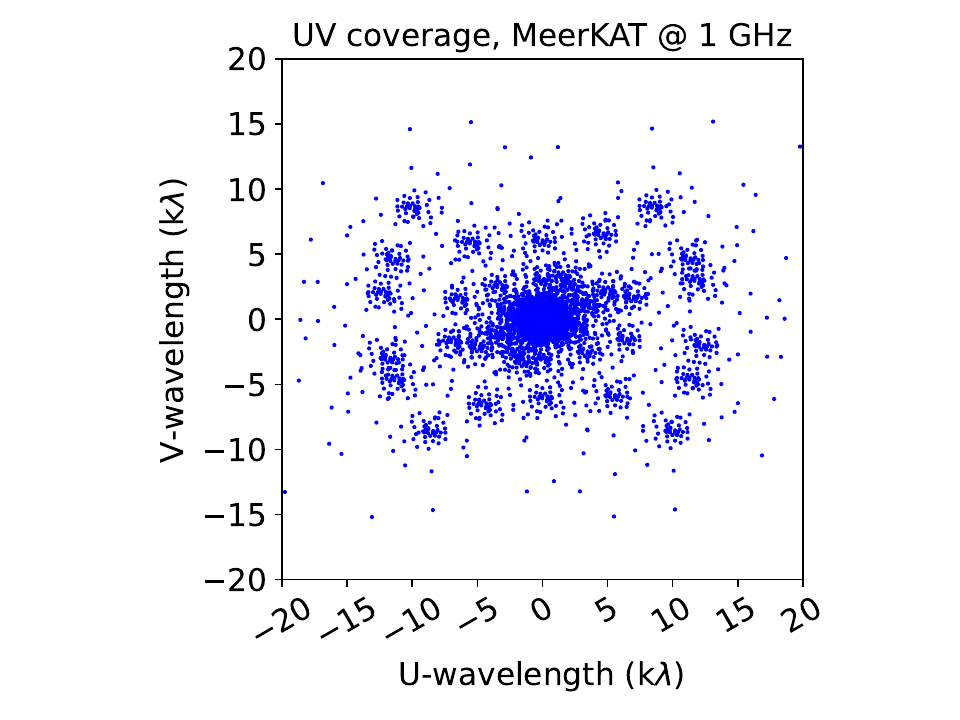}\includegraphics[trim={2.3cm 0cm 2cm 0cm},clip,scale=0.75]{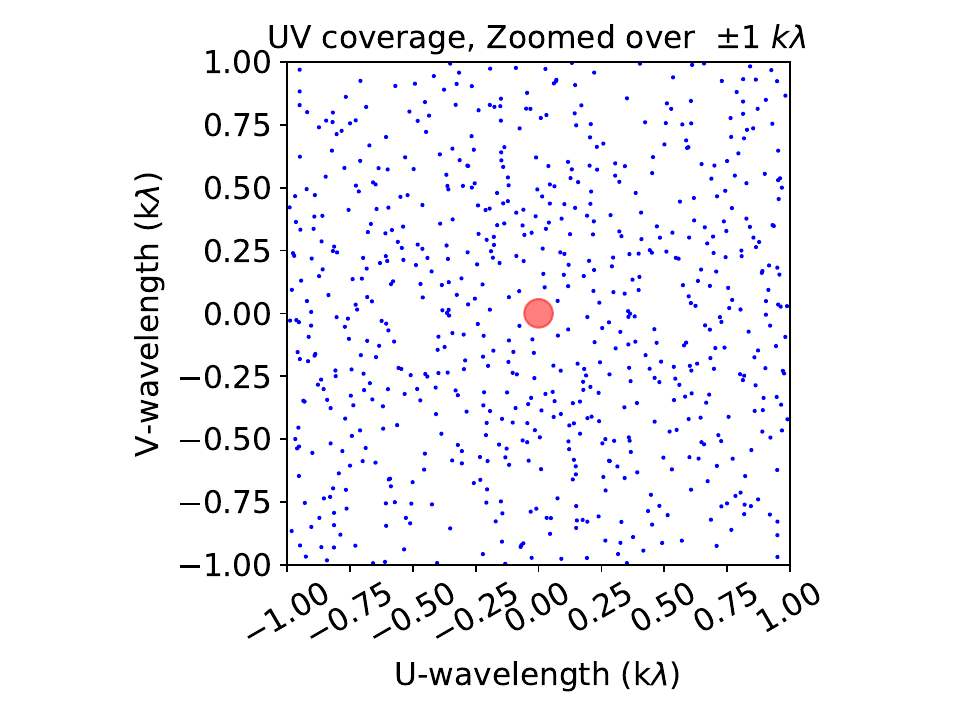}
\caption{\textbf{MeerKAT spectroscopic snapshot {\it uv-}coverage. Left panel: } Spectroscopic snapshot {\it uv}-coverage of MeerKAT at 1 GHz. \textbf{Right panel: }Same {\it uv}-coverage zoomed in over the central $\pm1 k\lambda$ region. The red circle at the origin corresponds to the {\it uv-}cell for a source with the size of the solar disc of 32 arcmin in angular scale. Note that {\it uv}-points corresponding to even the shortest baseline lie outside this  {\it uv}-cell.}
\label{fig:meerkat_uv_coverage}
\end{figure*}

MeerKAT is a new-generation instrument located in the MeerKAT National Park in the Northern Cape of South Africa. It consists of 64 dishes, each 13.5 m in diameter. Each MeerKAT dish is equipped with a cryogenically cooled receiver, making it extremely sensitive. At present, MeerKAT has three observing bands -- UHF (580 -- 1015 MHz), L (900 -- 1670 MHz), and S (1750 -- 3500 MHz) bands. The array is centrally condensed with about 39 dishes lying within 1~km and the remaining dishes distributed within a radius of $\sim$8 km. This provides MeerKAT with extremely good surface brightness sensitivity and also allows the generation of radio images with extremely high dynamic range (DR) and image fidelity \citep[e.g.][]{heywood2022}. The dense array layout of MeerKAT also implies that it has an excellent spectroscopic snapshot sampling in the Fourier plane ({\it uv}-plane) as shown in Figure \ref{fig:meerkat_uv_coverage}. This leads to a very well-behaved point-spread-function (PSF), making MeerKAT well-suited for high DR spectroscopic snapshot imaging. This capability is extremely useful for solar studies at radio wavelengths due to the rapid dynamics seen in solar radio emissions both along the spectral and temporal dimensions \citep{Oberoi2023,nindos2021}. There are several avenues where high DR snapshot imaging can lead to extraordinary science, ranging from the direct estimation of the magnetic field of the coronal mass ejections (CMEs) close to the Sun to studies of nonthermal emissions from extremely weak radio transients. 

MeerKAT is a precursor instrument to the mid-frequency telescope of the upcoming Square Kilometre Array Observatory \citep[SKAO;][]{Dewdney2017, SKAO2021}, referred to as the SKA-Mid. Given its larger number of elements and longer baselines, the SKA-Mid promises to be an even more capable instrument than MeerKAT and is expected to complete its construction phase sometime in 2028. The potential of the SKAO telescopes for providing major new insights into diverse aspects of solar physics is well recognized \citep{Nakariakov2015_SKA,Nindos2019_SKA}. Enabling solar science with the MeerKAT is not only expected to be scientifically very rewarding in its own right, but it also lies on the critical path to enabling solar science with the SKA-Mid and forms a part of the motivation for this work. A similar approach with the SKA-Low precursor and pathfinder, MWA, and LOFAR, have yielded very rich solar science dividends \citep[e.g.][]{Oberoi2023}. 

\begin{figure*}
    \centering
    \includegraphics[trim={2cm 2cm 0cm 0cm},clip,scale=0.4]{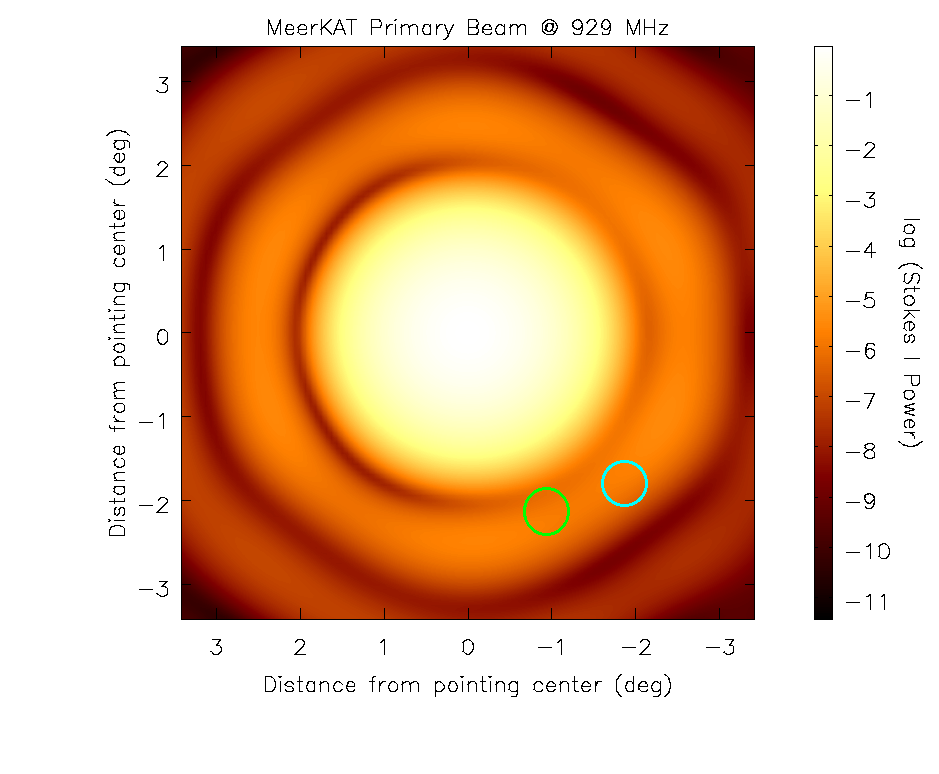}\includegraphics[trim={2cm 2cm 0cm 0cm},clip,scale=0.4]{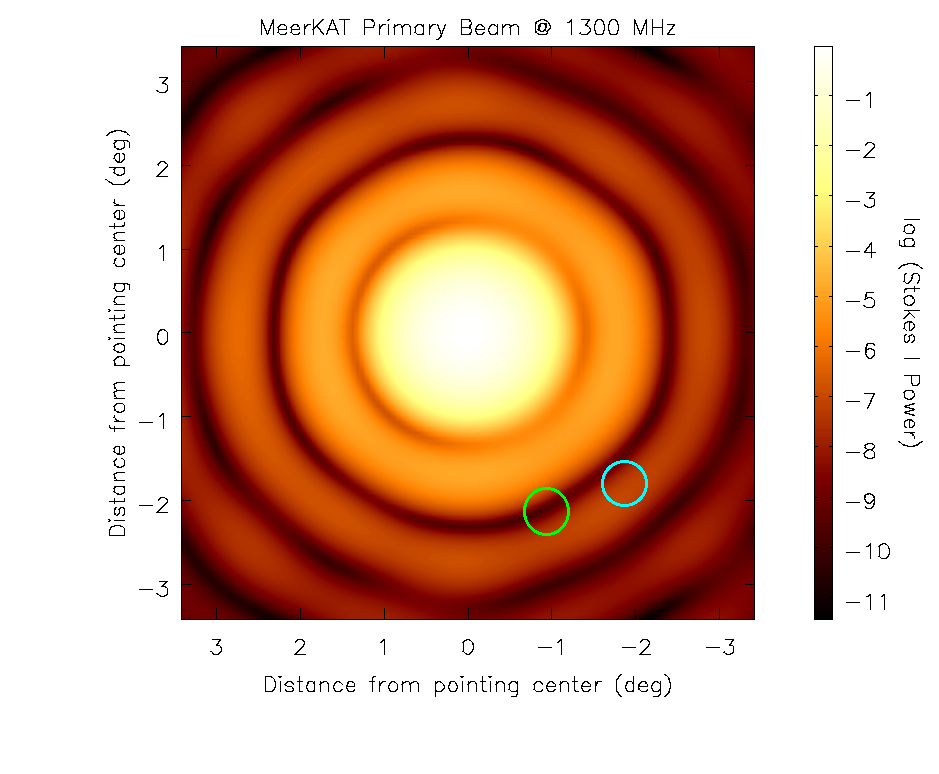}\\
    \includegraphics[trim={2cm 2cm 0cm 0cm},clip,scale=0.4]{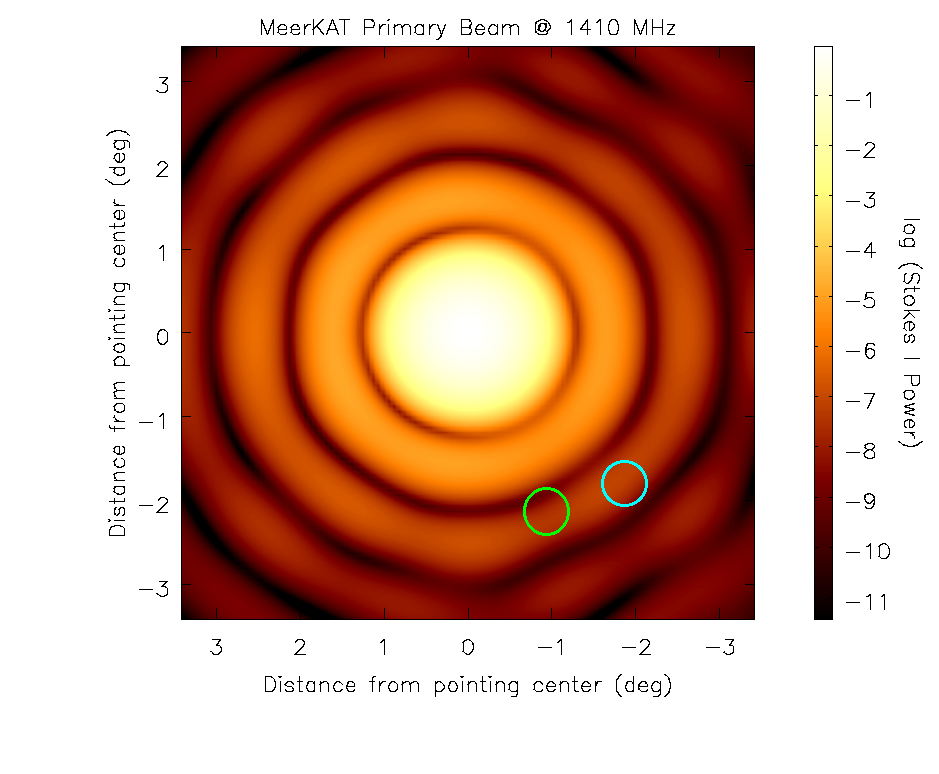}\includegraphics[trim={2cm 2cm 0cm 0cm},clip,scale=0.4]{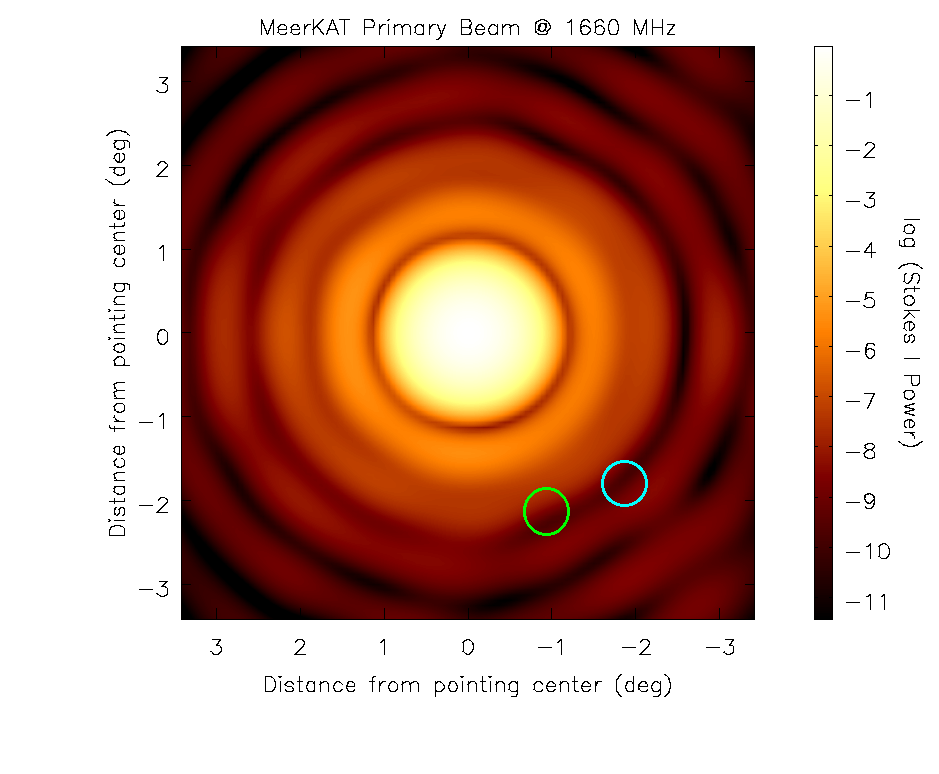}
    \caption{\textbf{Locations of the Sun in the primary beam.} Four different panels show MeerKAT holographic measured primary beam \citep{deVilliers_2022,deVilliers2023} at frequencies ranging from 929 MHz to 1660 MHz in increasing order. The cyan circle represents the location of the Sun on 26 September 2020 and the green circle represents that on 27 September 2020 for a particular observing scan. The position of the Sun changed azimuthally between different scans but lies at similar distances from the center of the primary beam.}
    \label{fig:sun_pos}
\end{figure*}

There have been previous attempts to image the Sun with MeerKAT. While solar images were generated successfully, these efforts were driven primarily by technological demonstration objectives and remain unpublished. Here we present the first detailed spectroscopic imaging study of the Sun with MeerKAT. Unlike standard astronomical observations, solar observations with any radio telescope pose several challenges. These challenges need to be addressed before MeerKAT can be used for solar observations. The primary reason behind this is that MeerKAT was designed for observing faint astronomical sources. To observe the Sun, the source with the highest flux density in the sky, the solar signal needs to be attenuated by many orders of magnitude to ensure that the astronomical signal stays in the linear regime of the instrument. However, the same attenuation cannot be used to observe the available calibrators as these sources are orders of magnitude weaker than the Sun. In the absence of such calibrator observations, it is hard to estimate the instrumental gains and efforts that are ongoing toward solving these issues. Hence, we have used a different strategy to observe and image the Sun. Instead of pointing at the Sun, we pointed $\sim$\,2.5$^{\circ}$ away to keep the Sun in the sidelobes of the primary beam to effectively attenuate the solar emissions. The sensitivity of MeerKAT is sufficient to image the Sun even when it is in the sidelobes of the primary beam. Availability of holographic measurements of the MeerKAT primary beam up to the second side lobes \citep{deVilliers_2022,deVilliers2023} allows us to obtain flux density calibrated solar images. We note that there are some shortcomings of this observing strategy. Among them, the non-uniform sensitivity over the solar disc due to the chromatic nature of the primary beam is the most important. Despite these shortcomings, our work substantiates the excellent spectroscopic snapshot imaging quality of MeerKAT solar data and showcases its potential for enabling excellent solar science. 

This paper is organized as follows. Section \ref{sec:observation} presents the details of the observations. Section \ref{sec:data_analysis} describes the data analysis procedure, including calibration, imaging, and primary beam correction. In Section \ref{sec:results}, we present our results and demonstrate some early results achieved using these data. Finally, in Section \ref{sec:conclusion}, we conclude by giving a future outlook of MeerKAT solar observations.

\begin{figure*}[!htpb]
    \centering
    \includegraphics[trim={0cm 1cm 2cm 0cm},clip,scale=0.14]{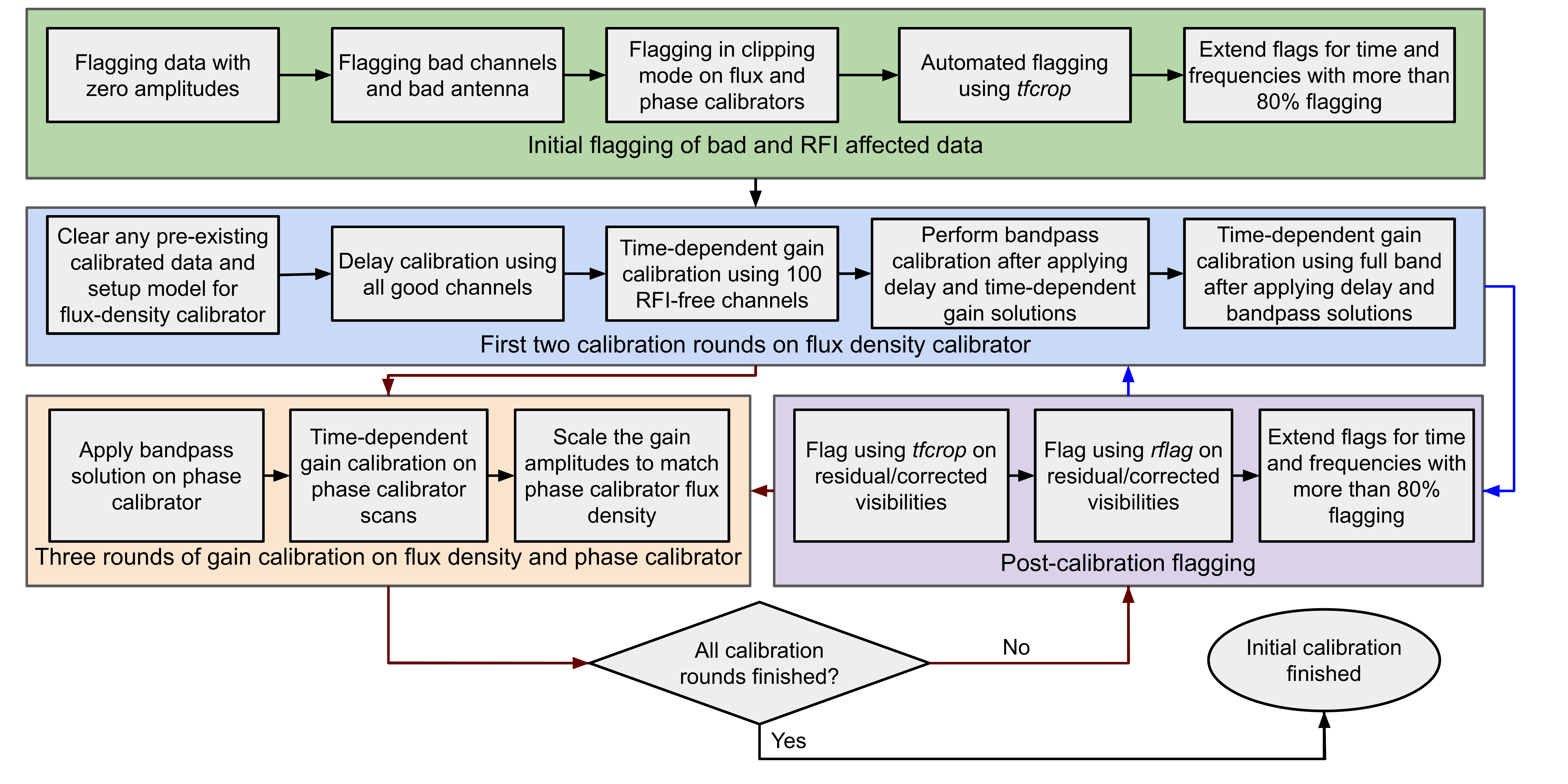}
    \caption{\textbf{Flowchart describing the flagging and initial calibration procedure.} Green box shows the steps of initial flagging on flux density and phase calibrators. The blue box shows the first two rounds of calibration steps on the flux density calibrator. The orange box shows the steps of the final three rounds of calibration on the phase calibrator. Each calibration round is followed by post-calibration flagging steps shown in the purple box. The first two cycles of calibration and post-calibration flagging are marked by blue arrows, and the last three rounds are marked by brown arrows.}
    \label{fig:cal_flowchart}
\end{figure*}
\begin{figure}[!htbp]
    \centering
    \includegraphics[trim={0cm 0cm -0.5cm 0cm},clip,scale=0.47]{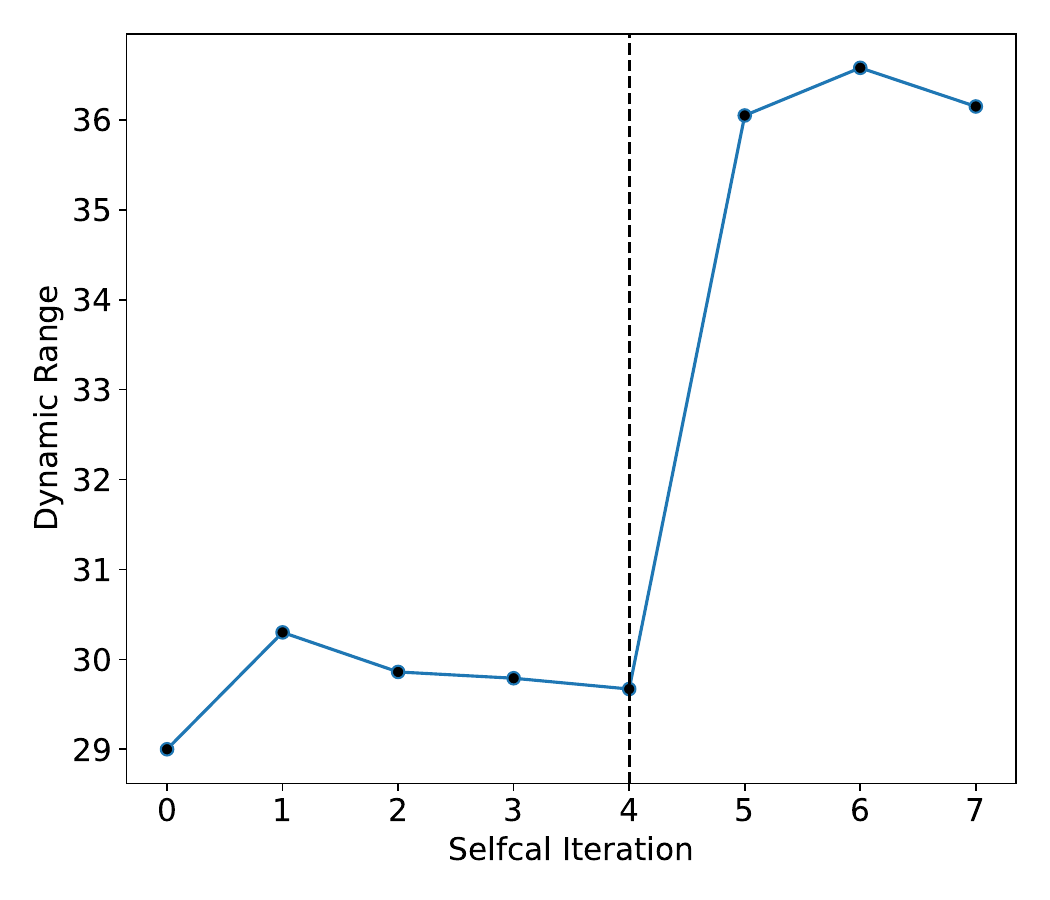}
    \caption{\textbf{Changes in the imaging dynamic range (DR) with self-calibration iterations for a single 20 MHz spectral slice centered at 929 MHz.} DR is estimated as the ratio of peak flux density and the rms noise close to the Sun. The black dashed line shows the iteration where amplitude-phase self-calibration is initiated.}
    \label{fig:selfcal}
\end{figure}
\begin{figure*}[!htpb]
    \centering
    \includegraphics[trim={2.5cm 14.5cm 2cm 1.6cm},clip,scale=0.98]{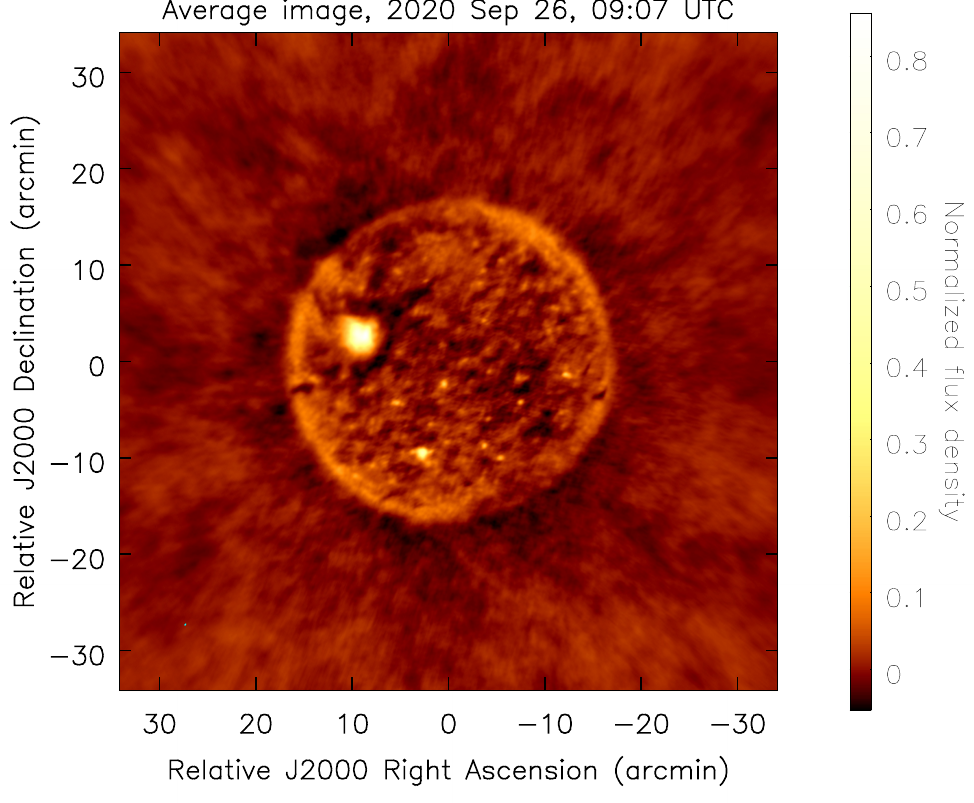}\\
    \includegraphics[trim={3cm 14cm 3cm 1cm},clip,scale=0.28]{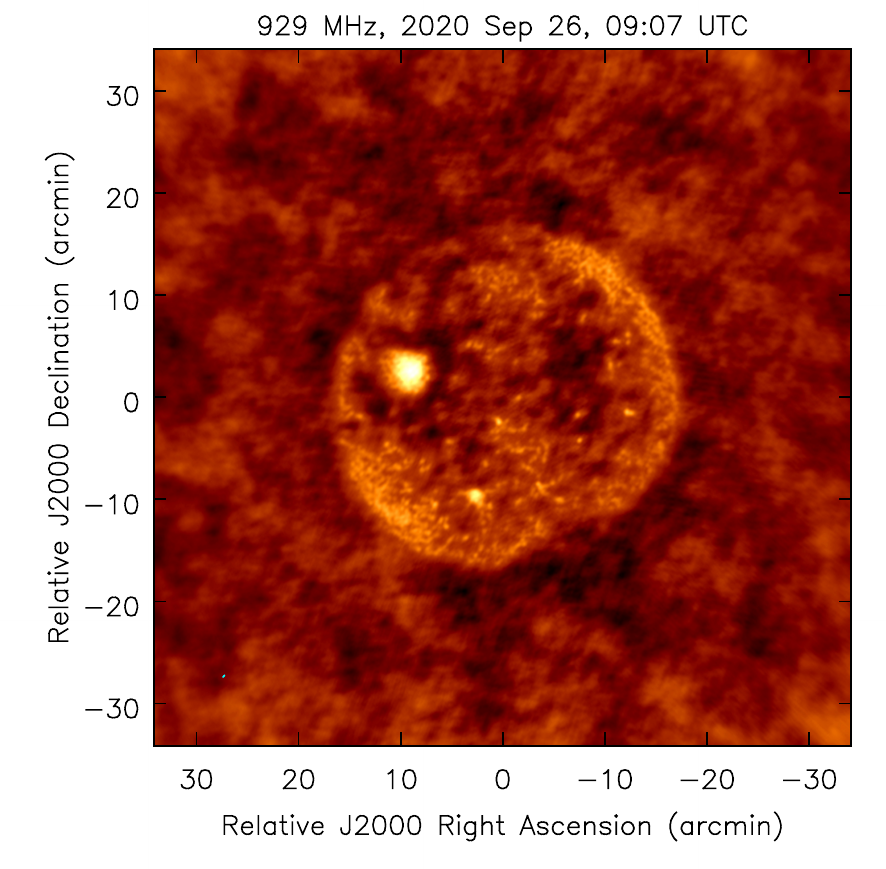}
    \includegraphics[trim={3cm 14cm 3cm 1cm},clip,scale=0.28]{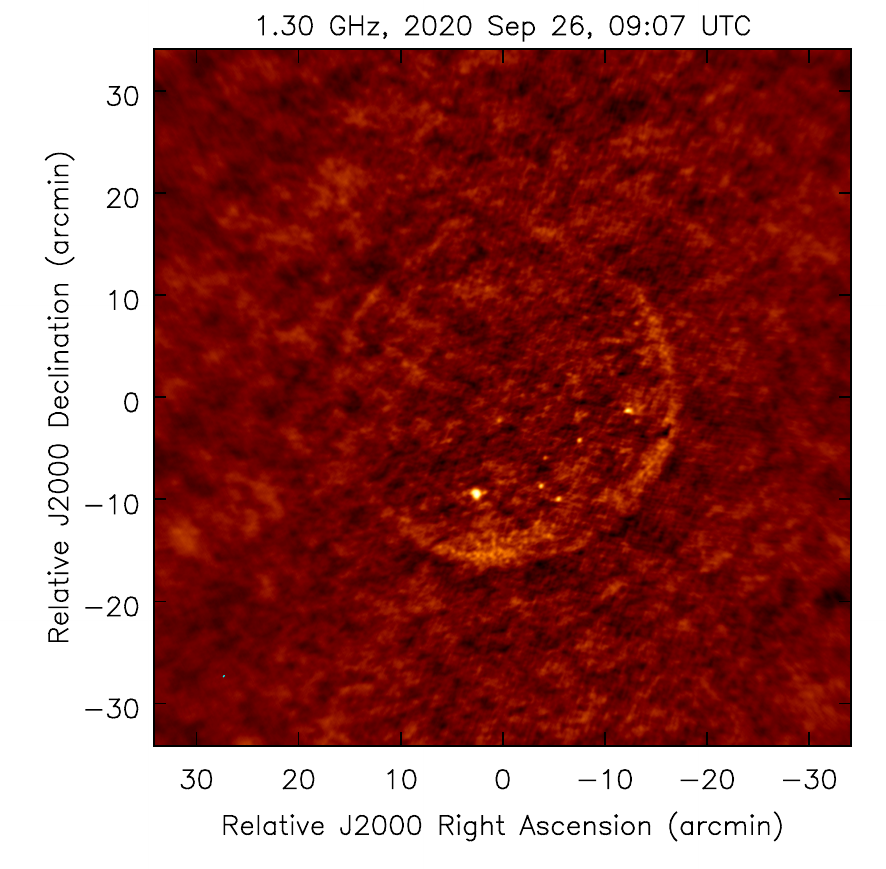}\includegraphics[trim={3cm 14cm 3cm 1cm},clip,scale=0.28]{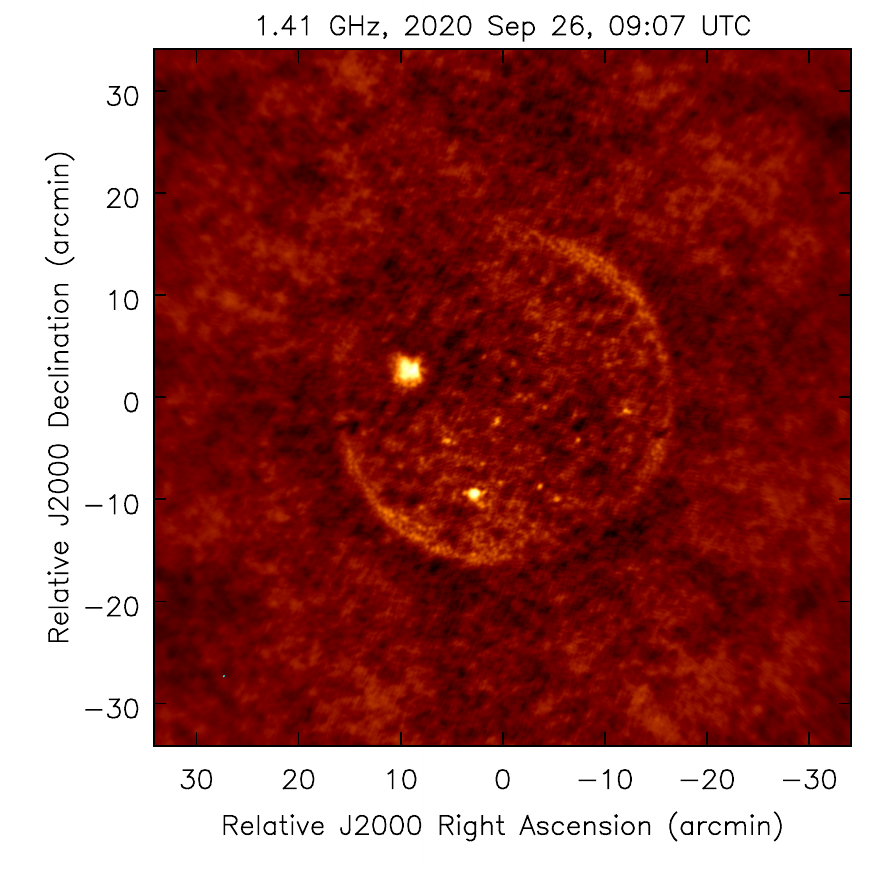}\includegraphics[trim={3cm 14cm 3cm 1cm},clip,scale=0.28]{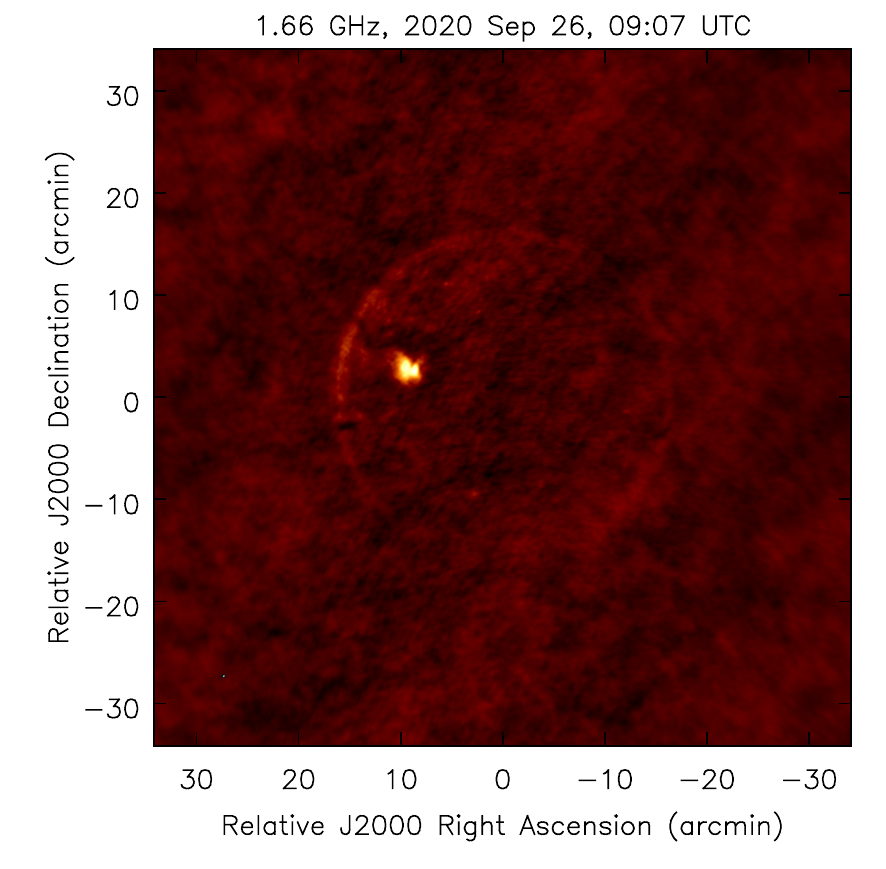}
    \caption{\textbf{Spectroscopic images of the Sun on 26 September 2020, 09:07 UTC. Top panel: }Normalized average image over the entire MeerKAT L-band. \textbf{Bottom panels:} Four sample images at different 20 MHz spectral chunks across the observing band. The small cyan dot at the bottom left is the PSF of the array.}
    \label{fig:spectral_images_26}
\end{figure*}

\begin{figure*}[!htpb]
    \centering
    \includegraphics[trim={0cm 0cm 0cm 0cm},clip,scale=0.85]{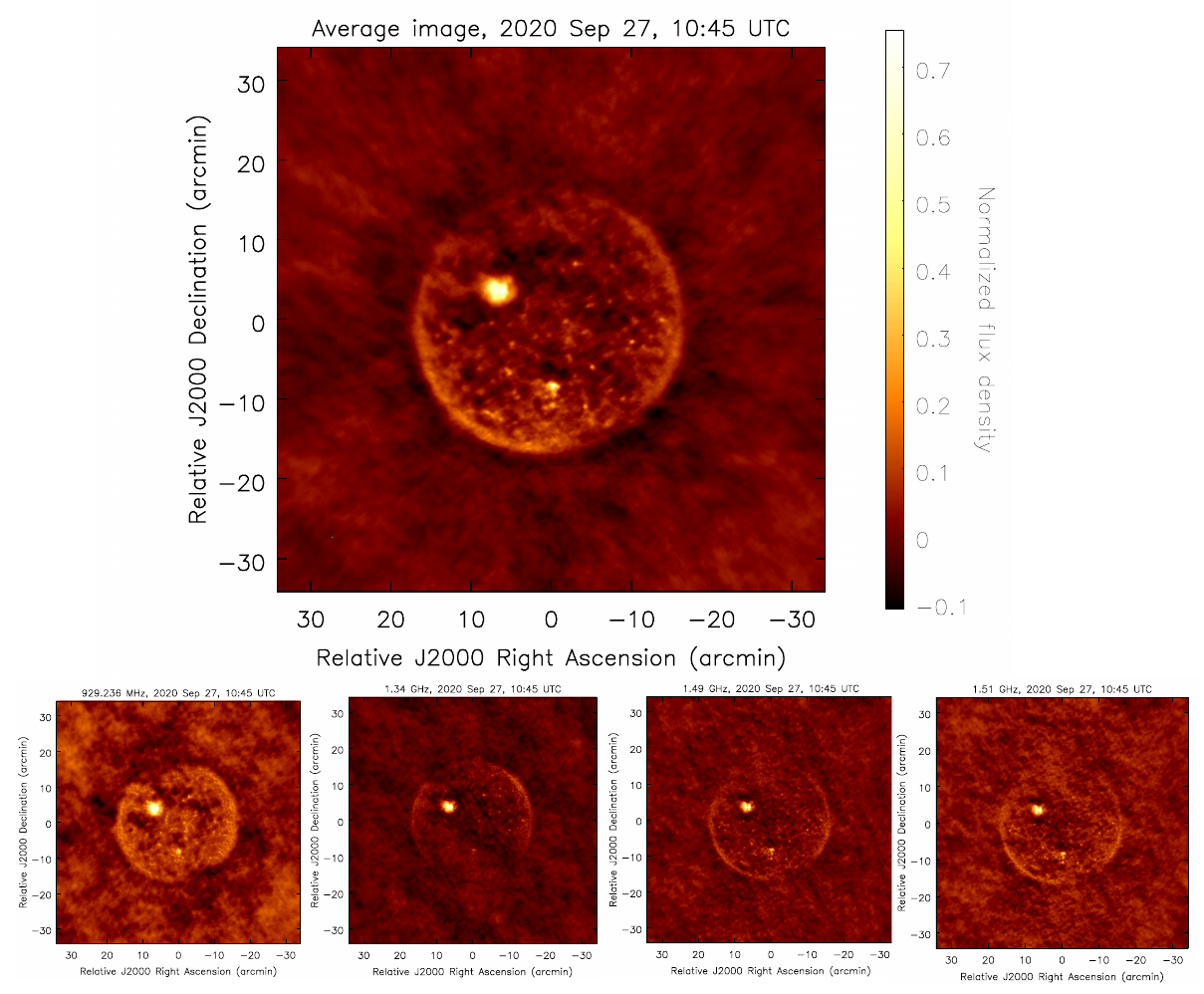}
    \caption{\textbf{Top panel:} normalized average image over the entire MeerKAT L band centered at the time on 2020 September 27, 10:45 UTC. \textbf{Bottom panels:} images at different 20 MHz spectral chunks across the L band. The small cyan dot at the bottom left represents the PSF of the array.}
    \label{fig:spectral_images_27}
\end{figure*}

\section{Observations}\label{sec:observation}
The observations reported here were done as a part of {\em Science Verification} (SSV) observation under the project ID SSV-20200709-SA-01. The raw visibilities for this project are available in the public domain through SARAO data archive\footnote{\url{https://archive.sarao.ac.za/}}. The observations were carried out during the 6th perihelion passage of Parker Solar Probe \citep[PSP,][]{psp2017} from 24 September 2020 to 30 September 2020. On each day, there are about 3 hours of observations centered around 10:30 UTC. In this paper, we present results from two of these epochs -- 26 September 2020 and 27 September 2020. 

Observations were done covering 856 -- 1712 MHz (usable frequency range 880 -- 1670 MHz) with 32 K spectral channels and 8 s temporal resolution. This provides us with data at about 26 kHz spectral resolution. Standard MeerKAT flux density calibrator, J0408-6545, was observed at the start of observations. J0408-6545 was used for bandpass and flux density calibration (hereafter referred to as fluxcal). J1239-1023 was used as a phase calibrator (hereafter referred to as phasecal) and observed between each consecutive solar scan. Since the Sun is a non-sidereal source, its RA -- DEC changes with time. Hence, the pointing center was changed every 15 minutes. For all the pointings, the Sun was kept $\sim$2.5$^\circ$ away from the pointing center. The position of the Sun in the primary beam for the particular scans from the two epochs analyzed here are shown for four different frequencies in Figure \ref{fig:sun_pos}. It turns out that at the lower part of the band, the Sun was in the first side lobe of the primary beam, while it was in the second side lobes or null at the higher parts of the band. This essentially makes the observations at the lower part of the observing band more sensitive than the high parts of the band. By keeping the Sun in the sidelobes of the primary beam, these observations can attenuate the solar signal by about $-30$ to $-90$ dB  (in power), depending on the frequency. This is essential to keep the total power levels in the linear regime all through the signal chain.

\section{Data Analysis}\label{sec:data_analysis}
Since the observation does not fall under the standard astronomical observation category, we did not use SARAO Science Data Processor (SDP) pipelines for the analysis. Instead, we did the analysis manually using Common Astronomy Software Application (\textsf{CASA}) \citep{mcmullin2007,CASA2022} for flagging and calibration and \textsf{WSClean} \citep{Offringa2014} for imaging.

\subsection{Flagging and Calibration}\label{subsec:flagging_calibration}
The flowchart for flagging and initial calibration procedures is shown in Figure \ref{fig:cal_flowchart}. Initial flagging is performed to remove bad antennas, bad frequency channels, and other strong radio frequency interferences (RFIs). Steps for initial flagging are marked by the green box in Figure \ref{fig:cal_flowchart}. After that, initial calibration rounds are done using the fluxcal and phasecal. A total of five rounds of initial calibration were done. The steps of the first two rounds of calibration on fluxcal are shown in the blue box of the same figure. After the first two rounds of initial calibrations on fluxcal, the next three rounds of calibrations are done on both fluxcal and phasecal, the steps of which are shown in the orange box in the same figure. Each calibration round is followed by post-calibration flagging steps marked by the purple box in the same figure. Detailed procedures for flagging and initial calibration steps are discussed in Appendix \ref{appendix_1}.

Once initial calibration and flagging are done, calibration solutions are applied to the solar scans, and self-calibration is performed. The Sun lies in the sidelobes of the MeerKAT primary beam, about 2.5$^{\circ}$ away from the nominal phase center at the center of the primary beam. To estimate and correct the complex gain towards the direction of the Sun, we first moved the phase center of the measured visibilities to the solar center before performing the self-calibration. Due to the chromatic nature of the primary beam, sensitivity across the solar disc varies with frequency. Also, being a non-sidereal source, the position of the Sun in the equatorial coordinate system changes with time. Hence, self-calibration is performed for each 20 MHz spectral chunk and 15-minute temporal chunk, separately. DR of the image is determined as the ratio of the peak flux density of the Sun and the rms noise close to the Sun. DRs are estimated from images before the primary beam corrections. Improvement in DR of the images of a 20 MHz spectral chunk centered at 929 MHz with self-calibration iterations is shown in Figure \ref{fig:selfcal}. To ensure the convergence of both phase-only and amplitude-phase self-calibration, we have compared DRs of three consecutive rounds. When the DR of three consecutive rounds changes by less than 1, we assume self-calibration has converged. A detailed description of key aspects of the self-calibration procedure followed here is presented in Appendix \ref{subsec:selfcal}.

\subsection{Final Imaging}\label{subsec:imaging}
Once the self-calibration is done, calibration solutions corresponding to the iteration with maximum DR are applied to the initial calibrated data. We then make final images of the Sun for each spectro-temporal chunk separately using all the baselines. All other imaging parameters -- the number of {\it w}-layers, visibility weighting, {\it uv}-taper, multiscale parameters, and pixel size are kept the same across the band. During the final imaging, we did not use any pre-defined mask. Instead, we use \textsf{auto-masking} feature available in \textsf{WSClean} to perform deconvolution down to 3$\sigma$, where $\sigma$ is the local rms. Due to the chromatic nature of the primary beam, the low-frequency part of the band has better sensitivity compared to the high-frequency part. At a single spectral slice, the sensitivity also varies across the solar disc due to variations of primary beam gain over the solar disc. This is evident from the spectral images shown in the bottom panels of Figure \ref{fig:spectral_images_26}. At higher frequencies, emissions from the brightest active region are detected with good detection significance, but the extended emission from the solar limb is not always detected at all frequency chunks.  At the same time, at 1.3 GHz even the brightest source corresponding to the active region is not detected, because it falls in the null between first and second sidelobes. This is evident from the cyan circle in the top right panel of Figure \ref{fig:sun_pos}. 
 
It is very interesting to compare the structures detected across the full spectral band with those seen in the simulated radio maps, described later in Section \ref{subsec:simulation}. The way this is generally done is by averaging in frequency. As we are imaging a very extended source in the sidelobes, the gain of the primary beam varies dramatically across the Sun and the chromaticity of the primary beam leads to large variations across frequency as well. Therefore correcting for the primary beam in individual spectral slices will lead to averaging of images with vastly varying spatio-spectral noise characteristics and can degrade the DR with which the emission features are detected. To avoid this issue, for morphological comparison, we have constructed the spectrally averaged image using images prior to primary beam correction. We have convolved all images at the resolution of the lowest frequency of the observing band. Then we normalized each 20 MHz spectral image with respect to the peak flux density and averaged all spectral chunks for a given scan to obtain a normalized full band image shown in the top panel of Figure \ref{fig:spectral_images_26}. This normalization ensures that the DR of individual spectral images is maintained while averaging and also ensures that we do not give undue weight to any specific spectral image at any location. DR of this normalized full-band image integrated over 15 minutes is $\sim500$, which is about an order of magnitude higher than the DR of individual 20 MHz spectral slices.

\subsection{Primary Beam Correction and Estimation of Absolute Flux Density}\label{subsec:pbcor_and_fluxcal}
At lower parts of the band ($<$1300 MHz), the Sun was in the first side lobe of the primary beam, while at the higher frequencies, it was in the second or higher sidelobes, as evident from Figure \ref{fig:sun_pos}. As the Sun is an extended source the primary beam response also varies across the solar disc. Hence, to obtain the absolute flux density, the direction and frequency dependence of the primary beam need to be corrected.

Holographic measurements of MeerKAT primary beam \citep{deVilliers_2022,deVilliers2023} at L-band are available\footnote{\href{https://skaafrica.atlassian.net/wiki/spaces/ESDKB/pages/1481572357/The+MeerKAT+primary+beam\#A-note-on-sidelobes}{MeerKAT holographic measurements of the primary beam}} over an extent of 4 degrees at an angular resolution of $\sim223$ arcsec. We did linear interpolation to obtain the beam values at each pixel of the image. For alt-az mount telescopes, the sky rotates with respect to the telescope beam, and the rotation angle is known as parallactic angle \citep{Meadows2007}. If the beam of the instrument is axially symmetric, then parallactic angle correction is not important for Stokes I imaging. As evident from Figure \ref{fig:sun_pos}, while the main lobe of the MeerKAT primary beam is close to axially symmetric, the same is not true for its sidelobes. In the present observation, the Sun was observed at the sidelobes of the primary beam. Hence, we rotate the primary beam by the parallactic angle before applying the primary beam corrections. We performed an image-plane-based primary beam correction using the array-averaged response. Being at the first/second side lobe of the primary beam, flux density measurements can have errors due to the uncertainty in primary beam measurements. Since the observations were done the sidelobes of the primary beam, antenna pointing errors can lead to larger than usual effects. Pointing errors arise when a beam is assumed to be steered precisely toward a certain direction, but in reality, it has a small offset from the desired direction. At present, MeerKAT has an rms pointing error of 0.64 arcmin, and no pointing calibration is done for MeerKAT science observations. This would lead to $\sim5\%$ error in primary beam power in the sidelobes \citep{deVilliers_2022}. Considering other kinds of errors (antenna to antenna variations in the side lobes of the primary beam, elevation-dependent effects due to gravity, etc. including the pointing jitters) as discussed in \cite{deVilliers2023}, we consider a conservative 10\% error on the absolute solar flux density measurements.

To obtain the absolute flux density of the solar emissions, we performed corrections of chromatic primary beam response for each 20 MHz spectral and 15-minute temporal slice individually. Since the primary beam measurements have larger uncertainties at low primary beam gain regions, for further spectroscopic analysis we have estimated absolute flux density for the spectral points which satisfy the following two conditions:
\begin{enumerate}
    \item The Sun should not lie beyond the first side lobe of the primary beam, and the value of the primary beam value towards the Sun should be $>$ 0.001 of the peak.
    \item The emission should be detected at a level $>$ 5$\sigma$, where $\sigma$ is the rms noise of the primary beam corrected image measured very close to the Sun. 
\end{enumerate}

\section{Results}\label{sec:results}
In this section, we present the results from spectroscopic solar imaging with MeerKAT and compare them with simulated MeerKAT solar maps at frequencies spanning our observations at L-band.

\subsection{Spectroscopic Solar Images using MeerKAT}\label{subsec:first_image}
\begin{figure*}[!htpb]
    \includegraphics[trim={1cm 0cm 3cm 0.5cm},clip,scale=0.7]{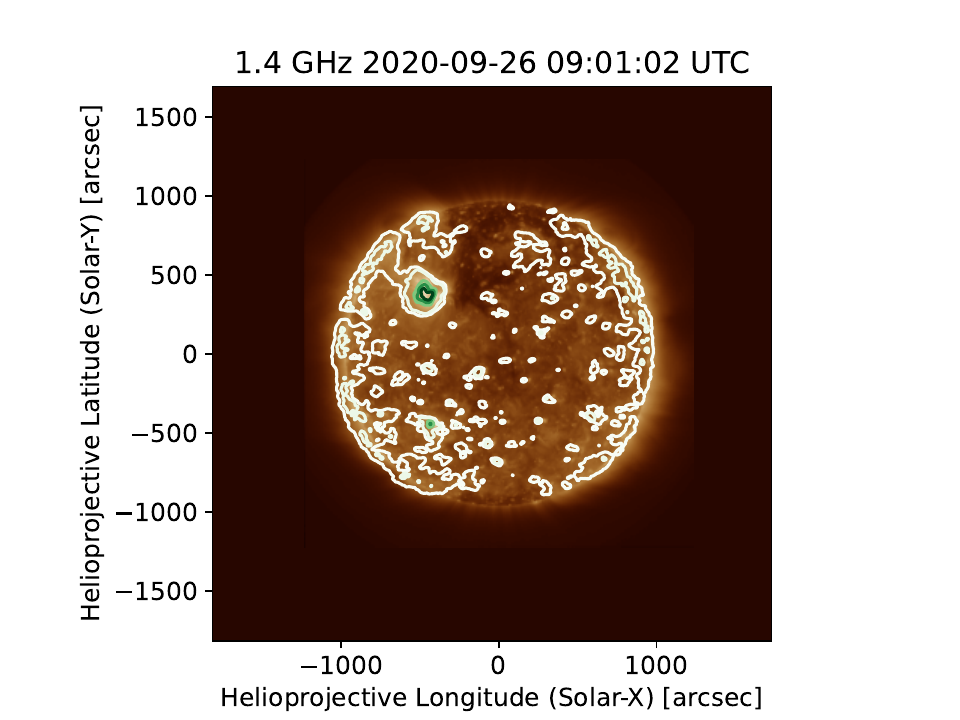}\includegraphics[trim={1cm 0cm 3cm 0.5cm},clip,scale=0.7]{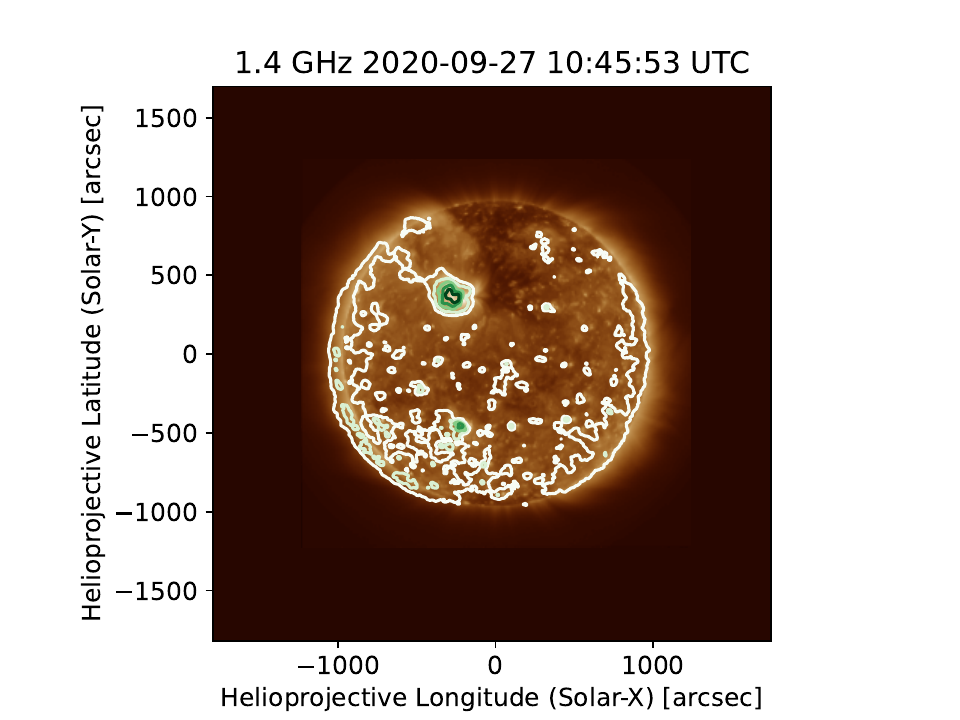}
    \caption{\textbf{MeerKAT solar images overlaid on AIA/SDO 193\AA\ images. Left panel: } Image for 26 September 2020, 09:07 UTC. Contours are at 4, 10, 40, 60, and 80\% of the peak flux density. \textbf{Right panel: }Image for 27 September 2020, 10:45 UTC. Contours are at 7, 20, 40, 60, and 80\% of the peak flux density. In both images, there are no noise peaks at the lowest contour level over a region $\sim1\deg\times1\deg$. The lowest contours in both images are chosen at 20$\sigma$ level, where $\sigma$ is measured rms close to the Sun.}
    \label{fig:aia_overlay}
\end{figure*}

The spectroscopic images of the Sun made using MeerKAT L-band observations on 2020 September 26 and 2020 September 27 are shown in Figures \ref{fig:spectral_images_26} and \ref{fig:spectral_images_27}, respectively. The top panel shows the averaged image over the entire MeerKAT L-band following the procedure described in Section \ref{subsec:imaging} and the lower panels show four sample spectroscopic images at individual 20 MHz spectral bands spanning the full observing band. The entire solar disc is visible once images over the full band are stacked together. We find that the solar disc is detected at $\sim50\sigma$ detection, where $\sigma$ is the rms noise close to the Sun. The diameter of the solar disc is found to be $\sim$35 arcmin, which is slightly larger than the optical disc. MeerKAT images are overlaid on 193\AA\ images from the Atmospheric Imaging Assembly (AIA) onboard Solar Dynamics Observatory \citep[SDO;][]{Lemen2012} in Figure \ref{fig:aia_overlay}. The largest active region is co-located with the brightest radio source in MeerKAT images. There are multiple small bright points visible in the AIA image, which are also detected in MeerKAT images with high significance. In both of these images, the diffuse quiet Sun emission from both limbs is also detected with good significance. Although visually both the images show features similar to those seen in the AIA images, we go further to verify this via a comparison with the corresponding simulated solar radio images.

\begin{figure*}[!htbp]
\centering
\includegraphics[trim={0cm 0cm 0cm -0.5cm},clip,scale=0.58]{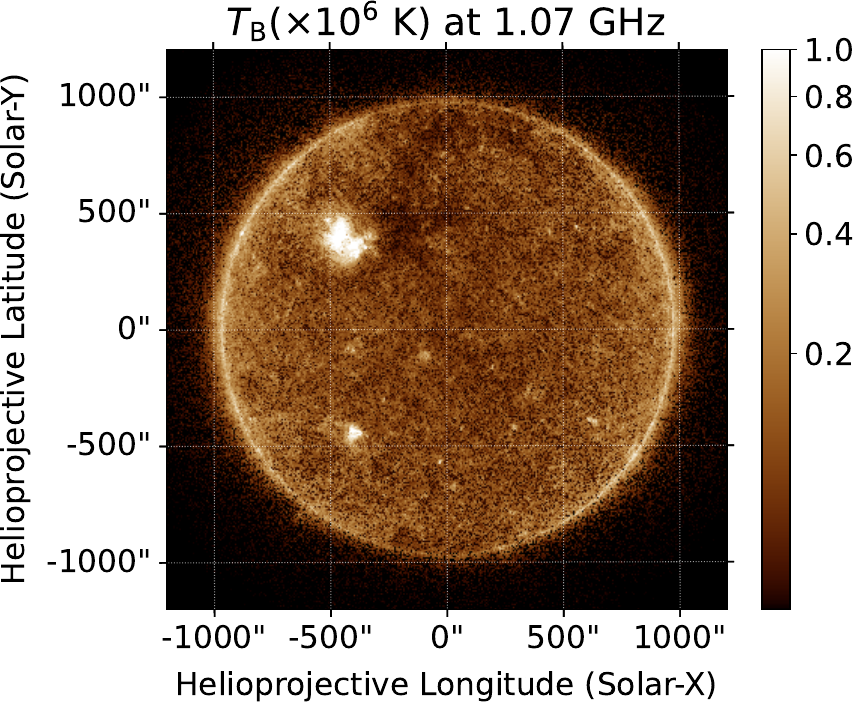}\includegraphics[trim={0cm 0cm 0cm -0.5cm},clip,scale=0.58]{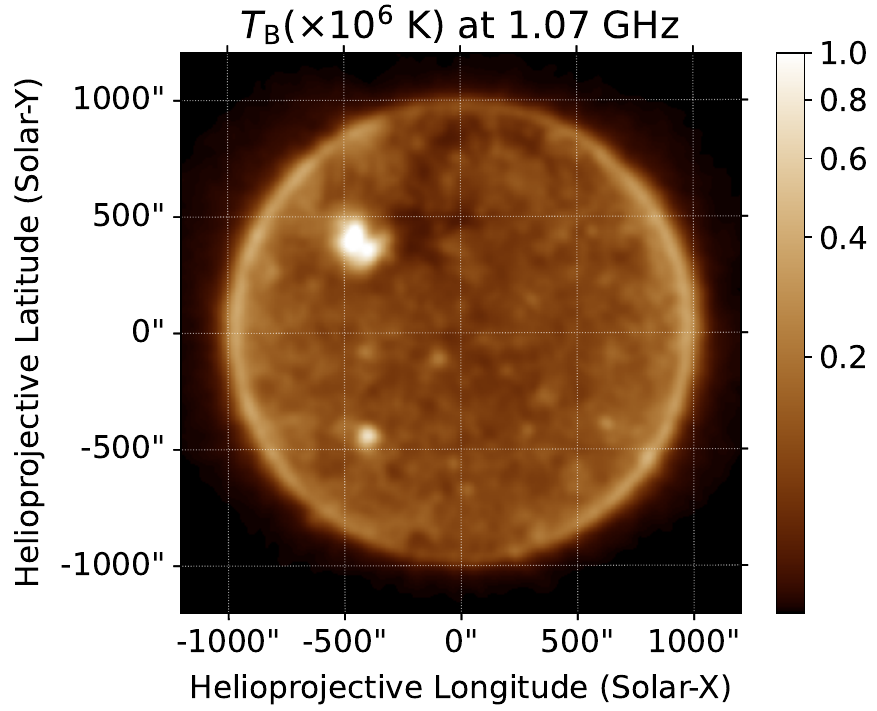}
\caption{\textbf{Simulated radio map on 26 September 2020, 09:06 UTC. Left panel: } Simulated radio map of the Sun at 1.07 GHz at a pixel scale of 4.8 arcsec. \textbf{Right panel: }Same image is convolved with the PSF of the observation at 1 GHz ($\sim$8arcsec).}
\label{fig:simulated_map}
\end{figure*}

\begin{figure*}[!htbp]
\centering
\includegraphics[trim={0cm 0cm 0cm 0cm},clip,scale=0.36]{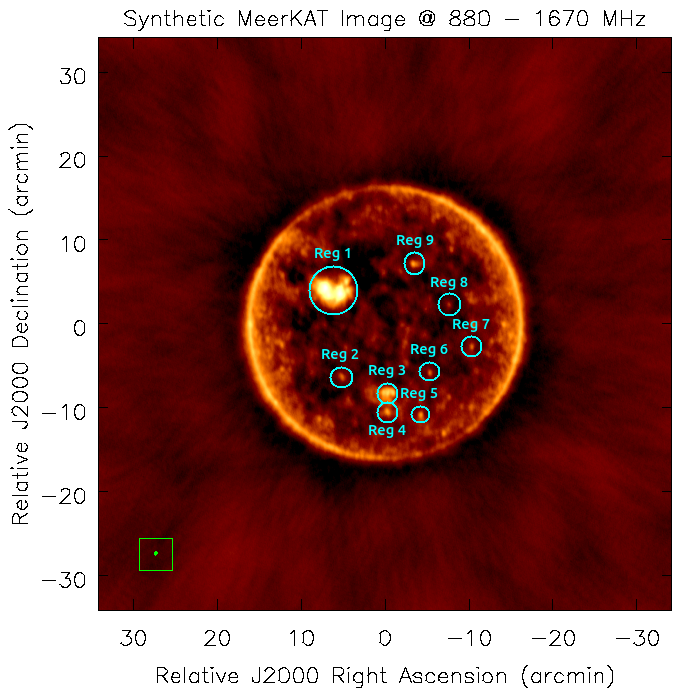}\includegraphics[trim={0cm 0cm 0cm 0cm},clip,scale=0.36]{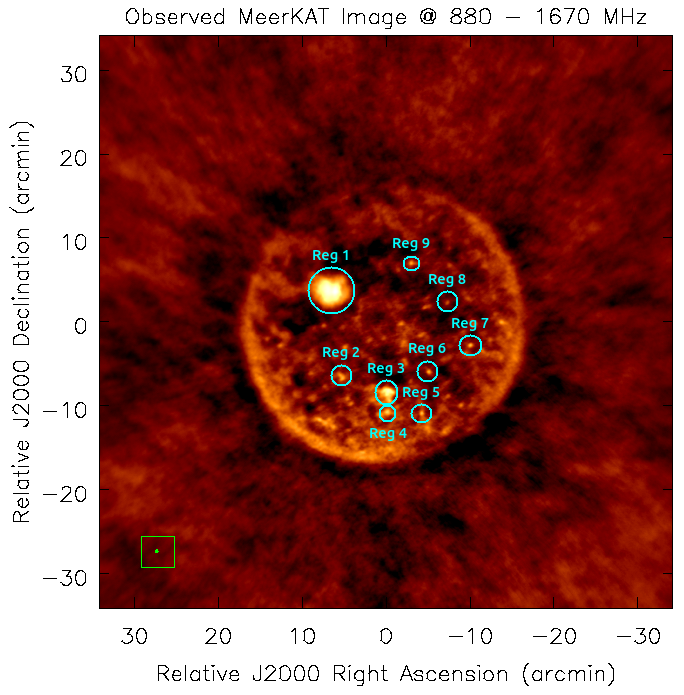}
\caption{\textbf{Comparison between synthetic and observed MeerKAT radio images on 27 September 2020, 10:45 UTC. Left panel: } Synthetic MeerKAT solar radio image. \textbf{Right panel:} Observed MeerKAT solar radio image. Both images are made using the entire frequency range from 880 -- 1670 MHz. In both the images multiple bright regions have been detected. Some of them are marked by cyan circles. Small green-filled circles at the bottom left corner marked by a green box is the PSF of the images.}
\label{fig:compare_sim_obs}
\end{figure*}

\subsection{Simulating Solar Radio Images and Spectra}\label{subsec:simulation}
The simulated images only aim to capture the thermal free-free emission. To generate simulated images, a differential emission measure (DEM) inversion is performed using images at different extreme ultraviolet wavelengths from the AIA/SDO. To reduce the computation time and improve the signal-to-noise of the obtained DEMs, the AIA images were smoothed to a resolution of $4.8$ arcsec, before DEM inversion. Though this degrades the resolution of these images, the resulting resolution is still finer than that of MeerKAT radio images. Following \cite{hannah2012,hannah2013}, we use the output of publicly available code\footnote{\url{https://github.com/ianan/demreg/tree/master/python}} to compute the expected free-free emission using the code developed by \citet{fleishman2021}. This simulation explicitly considers free-free optical depth of multi-thermal plasma while performing the radiative transfer calculations. A uniform line-of-sight depth of 100 Mm is assumed through the image. A chromospheric contribution has also been included, assuming that it is proportional to observations at 304\AA$\,$. The proportionality constant is determined assuming that the chromospheric contribution to the total brightness temperature is 10880 K \citep{zirin1991}. 

The left panel of Figure \ref{fig:simulated_map} shows the simulated $T_\mathrm{B}$ map of the Sun at 1.07 GHz and the right panel shows the same map smoothed at MeerKAT angular resolution. It is evident from these figures that there are emissions at a range of angular scales from instrumental resolution to the size of the solar disc. We note that the simulation does not incorporate any propagation effects like scattering or refraction. While their importance is well established, taking these into account appropriately is beyond the scope of this work.
\begin{figure*}[!ht]
    \centering
    \includegraphics[scale=0.4]{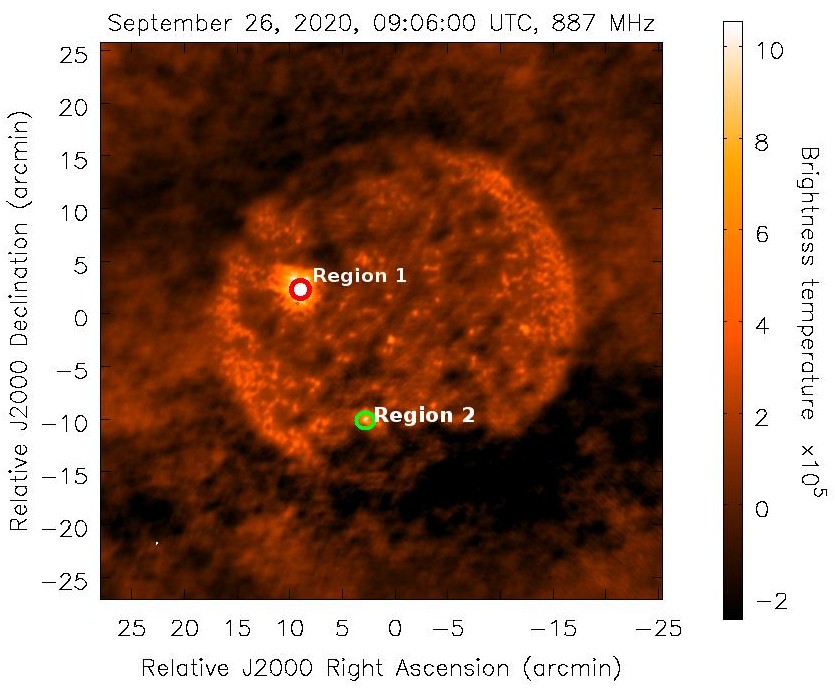}\\
    \includegraphics[scale=0.28]{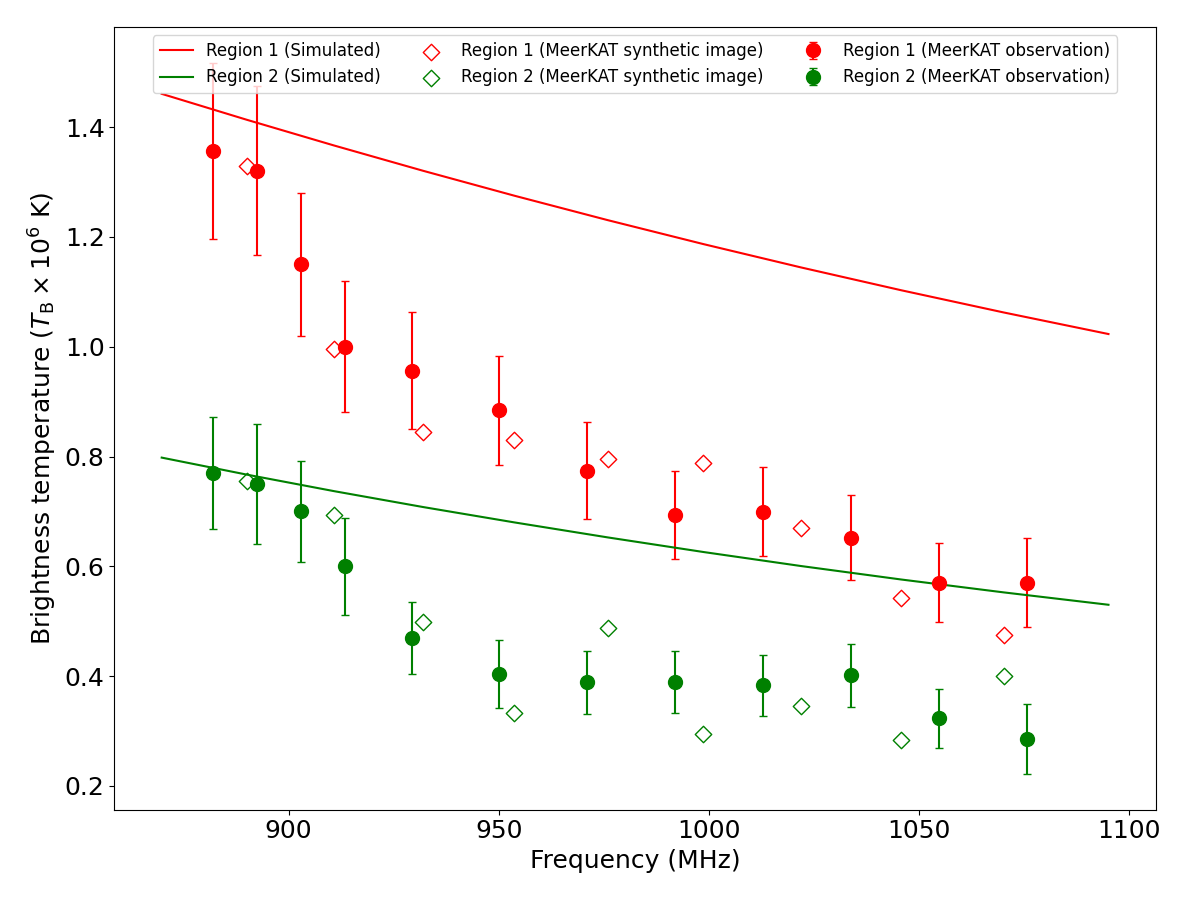}\includegraphics[scale=0.28]{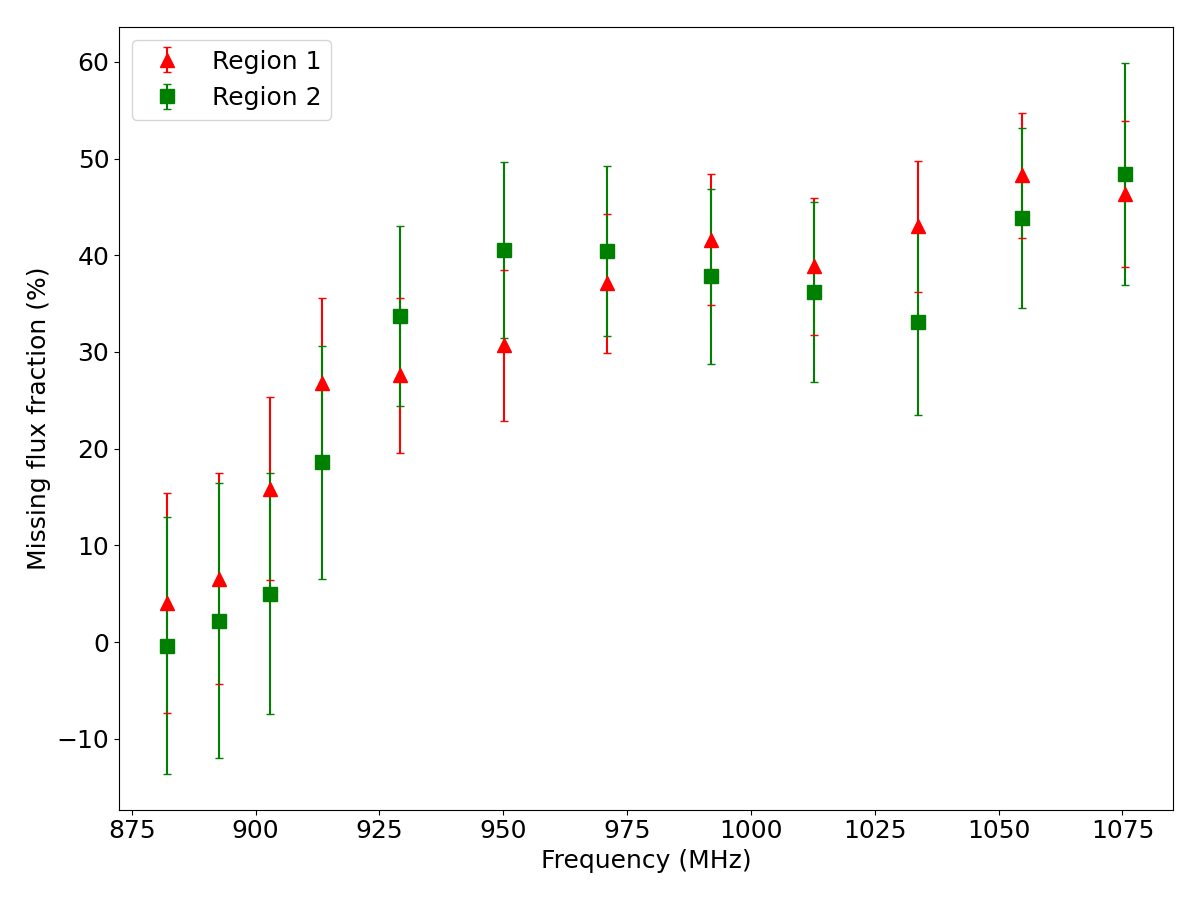}
    \caption{\textbf{Comparison of observed spectra with simulated spectra. Top panel: }A sample observed image at 887 MHz. Two regions are marked by red and green circles where spectra have been extracted. Spectra are extracted over a 20 arcsecond region centered around these regions. \textbf{Bottom left panel: }Spectra for regions 1 and 2 are shown in red and green colors, respectively. Solid lines represent the simulated spectra considering thermal emission (one sample image is shown in Figure \ref{fig:simulated_map}). Unfilled diamonds represent spectra from synthetic MeerKAT maps obtained from simulation. Filled circles represent measured spectra from MeerKAT observation on 26 September 2020. \textbf{Bottom right panel: }Missing flux fraction is shown as a function of frequency.}
    \label{fig:spectra}
\end{figure*}

\subsection{Comparing Simulated MeerKAT Images and Observations}\label{subsec:compare_simulation}
Radio interferometry is a Fourier imaging technique, where each baseline of the interferometer measures one Fourier component of the radio sky. Hence, the quality of the images and the scales of emission captured rely crucially on the sampling of the Fourier plane achieved by the interferometric observations. To build the appropriate simulated image for comparison with the observed MeerKAT images, we first create simulated visibilities from the simulated images using MeerKAT array configuration and observing parameters used for these observations. These visibilities are then inverted to make the synthetic MeerKAT radio image which would have been observed by MeerKAT.

A comparison between the synthetic MeerKAT image thus obtained and the observed MeerKAT image at the same time is shown in Figure \ref{fig:compare_sim_obs}. The left panel shows the synthetic MeerKAT map and the right panel shows the observed map from MeerKAT. The similarities between the simulated and observed images are very evident. The most striking similarities are the locations and relative intensities of the various bright points, some of them have been marked by cyan circles in both the panels of the same figure. There are also differences, the prominent ones are the presence of noise in the regions beyond the Sun, the limb being not as bright and well defined in MeerKAT image as compared to the synthetic image, and the differences in details of the morphology of the brightest active region. While the first of these can be attributed to the combined effect of the thermal noise associated with the image; the imperfections in the calibration and imaging process along with scattering in the solar atmosphere, may also play a role. 

\subsection{Comparison of Observed and Simulated Spectra}\label{subsec:obs_spectra}
In this section, we compare observed spectra with the expected spectra from the simulated images.
Absolute flux density calibration is done following the method described in Section \ref{subsec:pbcor_and_fluxcal} before extracting the spectra. We have extracted spectra for two bright active regions present on the Sun, which are marked by red and green circles in the top panel of Figure \ref{fig:spectra}, which is in the unit of brightness temperature. Corresponding brightness temperature spectra from these MeerKAT images for regions 1 and 2 are shown by the filled circles in the bottom left panel of Figure \ref{fig:spectra}. The criteria for points to be used for estimating absolute flux density were laid down in Section \ref{subsec:pbcor_and_fluxcal} and are met only below 1070 MHz. This limits the span of the spectra shown here.

We have extracted spectra of these two regions from the corresponding simulated radio maps (one sample simulated spectral map is shown in the left panel of Figure \ref{fig:simulated_map}), which are shown by solid lines in the left bottom panel of Figure \ref{fig:spectra}. Observed values shown by filled circles in the same figure are significantly different from the simulated values. We note that the simulation describes a rather ideal situation and can differ from observations due to several reasons, including the following: 
\begin{enumerate}
    \item Simulation assumes the thermal free-free emission from the coronal plasma to be the only emission mechanism in operation. In reality, however, the emission would be a superposition of the thermal free-free emission and gyrosynchrotron/gyroresonance emission at these frequencies \citep{Nindos2020}.
    \item The simulation ignores any propagation effects, while in reality refraction and scattering can lead to discernible effects.
    \item Interferometers are sensitive only to the brightness distributions at certain angular scales and not to a constant background. This implies that interferometers have a tendency to not be sensitive to emissions at large angular scales. The details of the largest angular scale to which an array is sensitive depend upon the details of the array configuration and the sampling of the Fourier domain achieved by the observation under study. This can lead to a reduction in the observed flux density when compared to the simulated values.
\end{enumerate}

It is feasible to isolate the impact of the last possibility mentioned above. To do this, we generate synthetic spectral maps of the Sun for MeerKAT array configuration as described in Section \ref{subsec:compare_simulation}, which can be compared directly with the observed MeerKAT solar maps for an apples-to-apples comparison. The spectra from these synthetic simulated maps are shown by unfilled diamonds in the bottom left panel of  Figure \ref{fig:spectra}. The spectra from synthetic maps are consistent with those from the observations. This demonstrates that the large discrepancy between the observed and simulated spectra (shown by solid lines in the same figure) is primarily due to the missing flux density in MeerKAT solar maps. The ratio of the flux density measured in the observed map to that in the simulated map is defined to be missing flux density fraction and is plotted in the bottom right panel of Figure \ref{fig:spectra}.  The red circle at the center of {\it uv}-plane shown in Figure \ref{fig:meerkat_uv_coverage} corresponds to the {\it uv}-cell for a source with the size of the solar disc of 32 arcmin in angular scale. There is no {\it uv}-sampling in that cell. Hence, it is expected to have missing flux for the Sun for the MeerKAT array configuration.

The missing flux density fraction decreases with the decrease in frequency. For a given array layout, one sample with increasingly shorter spacings in the {\it uv}-plane with decreasing frequency and missing flux density fraction at larger angular scales is expected to drop. The observed variation in the missing flux density fraction shows this trend and substantiates this to be the major cause of the observed differences between the simulated and observed MeerKAT solar spectra. While the other two reasons mentioned above could also be contributing to the observed differences, their effects, however, are smaller than the uncertainty on these measurements.

\section{Conclusion and Future Work}\label{sec:conclusion}
The Sun is an extremely complicated radio source with emissions at angular scales ranging from a few arcseconds to the size of the solar disc at GHz frequencies, as is evident from the simulated radio map shown in the left panel of Figure \ref{fig:simulated_map}. Solar emissions also show rapid spectro-temporal variations. Hence to study the solar radio emission at GHz frequencies, one requires a high DR and high-fidelity spectroscopic snapshot imaging of the Sun. Sufficiently dense spectroscopic snapshot {\it uv}-coverage of MeerKAT allows high DR and high-quality imaging of the Sun. Solar observation with MeerKAT has not yet been commissioned, and these observations were done keeping the Sun in the sidelobes of the primary beam. 

Here, we have presented the first detailed spectroscopic imaging study of the Sun with MeerKAT.  Given the well-behaved spectroscopic snapshot PSF and the precise calibration, the images presented here are the highest quality spectroscopic snapshot solar images at these frequencies available to date. To demonstrate the capability of MeerKAT in producing very high-quality spectroscopic solar images, we have compared MeerKAT images with the synthetic MeerKAT images designed to sample the same Fourier components as MeerKAT observations. The correspondence between the observed and simulated images shown in Figure \ref{fig:compare_sim_obs} is remarkable and it is evident from the fact that several weak solar emissions present in the synthetic image are detected with high significance in MeerKAT image as well. 

Although the spatial structures in the observed image match well with the simulated image, from a scientific perspective, it is also important to test the ability of MeerKAT to determine the flux densities and spectra of solar features. As substantiated in Section \ref{subsec:obs_spectra}, MeerKAT spectra show evidence of missing flux at higher frequencies which drops to insignificant levels by about 900 MHz. An implication is that while MeerKAT images in the UHF band are not expected to suffer from the missing flux density issue, one will need to be careful about the missing flux density at the L-band and higher. A comparison with different regions from the simulated maps radio maps might provide a good way to quantify the missing flux density fraction for specific observations.

While it is adequate for demonstrating the feasibility of MeerKAT for solar observations and evaluating the quality of the images it can deliver, a key limitation of the present observing approach arises from the issues related to imaging a source of large angular size in the chromatic primary beam sidelobes. Since the primary beam gain at the first side lobe is $\sim-30$ dB, we lose the sensitivity by a factor $\sim1000$ compared to the observation done using the main lobe of the primary beam. This was however necessitated by the requirement to attenuate the solar signals to a level that would keep the signal chain downstream in its linear regime. Hence, a preferable approach for solar observing will be to keep the Sun in the main lobe of the primary beam and adjust the gains of the appropriate elements of the signal chain to attenuate the signal to the required levels. Some members of this team are currently working with the MeerKAT engineering team to identify the suitable attenuation for solar observations and to develop a calibration strategy for solar observations performed along these lines. Once enabled, we are convinced that, with its high-quality spectroscopic snapshot solar imaging capability, MeerKAT solar observations will open a new frontier in solar radio physics.

\facilities{MeerKAT \citep{meerkat2016,Chen_meerkat2021}, Solar Dynamics Observatory \citep[SDO;][]{Lemen2012}.}

\software{astropy \citep{price2018astropy}, matplotlib \citep{Hunter:2007}, Numpy \citep{Harris2020}, CASA \citep{mcmullin2007,CASA2022}, WSClean \citep{Offringa2014}}

\begin{acknowledgments}
The MeerKAT telescope is operated by the South African Radio Astronomy Observatory, which is a facility of the National Research Foundation, an agency of the Department of Science and Innovation. The authors acknowledge the contribution of all those who designed and built the MeerKAT instrument. We also thank the anonymous referee for the comments and suggestions, which have helped improve the clarity and presentation of this work. We thank Sharmila Goedhart and Fernando Camilo for their comments on an earlier draft of the manuscript, which helped improve its technical correctness. We also thank the MeerKAT team for their assistance. D.K. and D.O. acknowledge the support of the Department of Atomic Energy, Government of India, under project no. 12-R\&D-TFR-5.02-0700. D.K. thanks Barnali Das (CSIRO, Australia) for the useful comments on the manuscript. This research has made use of NASA's Astrophysics Data System (ADS).    
\end{acknowledgments}

\appendix
Here we discuss the analysis steps followed in some detail, with emphasis on self-calibration.

\section{Initial Flagging and Calibration}\label{appendix_1}
We performed flagging and calibration using Common Astronomy Software Applications \citep[CASA;][]{CASA2022}. We have used J0408-6545 as the flux density calibrator for both the epochs and its model is described in MeerKAT calibration manual\footnote{\href{https://skaafrica.atlassian.net/wiki/spaces/ESDKB/pages/1452146701/L-band+gain+calibrators}{J0408-6545 model}}. The flagging and calibration of the flux-density and phase calibrators follow essentially the standard iterative procedure as indicated in the blocks marked by green and blue, respectively, in Figure \ref{fig:cal_flowchart}. For solar scans, only the data with zero amplitudes, known bad spectral channels and bad antennas were flagged. No automated flagging was performed on uncalibrated solar scans. 

Being a non-sidereal source, the equatorial coordinates of the Sun change with time. To accommodate this, we treated each 15-minute solar scan independently for self-calibration and imaging. We also did not consider the full spectral range simultaneously during self-calibration and imaging, because both the solar flux density and the primary beam gains in the sidelobes vary significantly with frequency. Hence, after applying the initial calibration solutions obtained towards the pointing center, we split every 15-minute solar scan into 20 MHz spectral chunks for self-calibration and imaging.

\section{Self-calibration}\label{subsec:selfcal}
Although the Sun is in the sidelobes of the primary beam, it is still the source with the highest flux density contributing to the observed visibilities. Before primary beam correction, the next brightest source in the field has a flux density of 38 mJy/beam, while the peak flux density on the Sun is about 1.7 Jy/beam. The total integrated flux density of background sources is about 0.5 Jy, whereas the integrated flux density of the Sun is about 15 Jy. Since the total contribution from background sources is only about 3.3\%, they do not affect the self-calibration significantly. 

As the Sun is present about 2.5$^\circ$ away from the pointing center, gain solutions towards the Sun may be different from those determined towards the pointing center. To estimate and correct the complex gains towards the direction of the Sun, we first shifted the phase center of the visibilities to the center of the Sun for each solar scan and then performed self-calibration on each spectral chunk of every scan separately. This has been done to account for the chromatic primary beam response and also the spectral variations in emissions from solar structures. 

Another major challenge in self-calibrating solar observation is the large variation in the flux distribution across baselines of varying lengths. Assuming the Sun to be a uniformly illuminated disc of size 32 arcmin, the first null of the visibility amplitude distribution lies close to 100$\lambda$, and the amplitudes of visibilities for baselines less than 100$\lambda$ increase very rapidly as one moves towards shorter baselines. Hence one needs good {\it uv}-coverage at baseline lengths $<100\lambda$ to properly model this emission. However, there are a limited number of short baselines of length $\leq100\lambda$ at MeerKAT for L-band observations. To avoid issues arising due to sparse {\it uv-}coverage at $\leq100\lambda$, we only use baselines $>100\lambda$ during the self-calibration. 

We followed the following self-calibration steps:
\begin{enumerate}
    \item First we make a circular mask of diameter 35 arcmin centered at the Sun.
    \item An image is made using \textsf{WSClean} from the calibrated data using antenna gain solutions obtained from flux density and phase calibrators. We choose baselines $>100\lambda$ and use \textsf{briggs} weighting \citep{Briggs1995} with robustness 0\footnote{\href{https://wsclean.readthedocs.io/en/latest/image_weighting.html}{Definition of robustness parameter as per WSClean}} along with a circular taper at 19k$\lambda$. 
    \item We keep {\it w}-stacking on and number of {\it w}-plane is chosen automatically by \textsf{WSClean}.
    \item Deconvolution is performed using the mask centered on the Sun. The average rms ($\sigma$) close to the Sun is about 0.1 Jy and we performed deconvolution down to 3$\sigma$, 0.3 Jy. We used multiscale deconvolution with Gaussian scale sizes 0, 5, 9, 15, 25, and 35 times the pixel size, where one pixel is chosen to be of 1 arcsec.
    \item Deconvolved model of the Sun is converted into model visibilities using \textsf{WSClean} and used for self-calibration.
    \item We performed four rounds of phase-only self-calibration followed by five rounds of amplitude-phase self-calibration. Time-dependent gain solutions are calculated using \textsf{CASA} task \textsf{gaincal} at 1-minute time interval using \textsf{solmode=L1R} and \textsf{minsnr=3} using baselines $>100\lambda$.
    \item Due to sidelobe response, visibility amplitudes for two parallel-hand polarizations ($XX$ and $YY$) could be different. Hence, during the amplitude-phase self-calibration, we make separate sky models for $XX$ and $YY$ polarizations. 
    \item Since some antennas, the ones with primarily long baselines, may not have sufficient signal-to-noise ratio for performing self-calibration, time-dependent gain solutions are applied using \textsf{applycal} task of \textsf{CASA} in \textsf{calonly} mode to retain the long baseline antennas with the initial calibration solutions. 
\end{enumerate}
We have calculated the DR of the spectral images as the ratio of peak solar flux density and the measured rms close to the Sun. Change in DR with self-calibration iterations is shown in Figure \ref{fig:selfcal}. When the DR of three consecutive self-calibration rounds does not change by more than 1, we consider the self-calibration to have converged. Once phase-only self-calibration has converged, we move to amplitude-phase self-calibration. We noticed there is an increase in DR by about 20\% when amplitude-phase self-calibration is initiated. Once the amplitude-phase self-calibration has also converged, the self-calibration loop is stopped. Though the improvements in DR with self-calibration iterations were comparatively modest, the final images shown in Figure \ref{fig:spectral_images_26} don't show any significant deconvolution artifacts implying that good calibration has been achieved.

\bibliography{sample631}{}

\begin{thebibliography}{}
\expandafter\ifx\csname natexlab\endcsname\relax\def\natexlab#1{#1}\fi
\providecommand{\url}[1]{\href{#1}{#1}}
\providecommand{\dodoi}[1]{doi:~\href{http://doi.org/#1}{\nolinkurl{#1}}}
\providecommand{\doeprint}[1]{\href{http://ascl.net/#1}{\nolinkurl{http://ascl.net/#1}}}
\providecommand{\doarXiv}[1]{\href{https://arxiv.org/abs/#1}{\nolinkurl{https://arxiv.org/abs/#1}}}

\bibitem[{Briand {et~al.}(2022)Briand, Cecconi, Chrysaphi, Girard, Grießmeier,
  Hariharan, Loh, Murphy, Kantepalli, Zarka, \& Zhang}]{Briand2022}
Briand, C., Cecconi, B., Chrysaphi, N., {et~al.} 2022, URSI Radio Science
  Letters, 4, \dodoi{10.46620/22-0017}

\bibitem[{{Briggs}(1995)}]{Briggs1995}
{Briggs}, D.~S. 1995, in American Astronomical Society Meeting Abstracts, Vol.
  187, American Astronomical Society Meeting Abstracts, 112.02

\bibitem[{Chen {et~al.}(2021)Chen, Barr, Karuppusamy, Kramer, \&
  Stappers}]{Chen_meerkat2021}
Chen, W., Barr, E., Karuppusamy, R., Kramer, M., \& Stappers, B. 2021, Journal
  of Astronomical Instrumentation, 10, 2150013,
  \dodoi{10.1142/S2251171721500136}

\bibitem[{{de Villiers}(2023)}]{deVilliers2023}
{de Villiers}, M.~S. 2023, Astronomical Journal, 165, 78,
  \dodoi{10.3847/1538-3881/acabc3}

\bibitem[{de~Villiers \& Cotton(2022)}]{deVilliers_2022}
de~Villiers, M.~S., \& Cotton, W.~D. 2022, The Astronomical Journal, 163, 135,
  \dodoi{10.3847/1538-3881/ac460a}

\bibitem[{Dewdney {et~al.}(2017)Dewdney, Braun, \& Turner}]{Dewdney2017}
Dewdney, P.~E., Braun, R., \& Turner, W. 2017, in 2017 XXXIInd General Assembly
  and Scientific Symposium of the International Union of Radio Science (URSI
  GASS), 1--4, \dodoi{10.23919/URSIGASS.2017.8105425}

\bibitem[{{Fleishman} {et~al.}(2021){Fleishman}, {Kuznetsov}, \&
  {Landi}}]{fleishman2021}
{Fleishman}, G.~D., {Kuznetsov}, A.~A., \& {Landi}, E. 2021, The Astrophysical
  Journal, 914, 52, \dodoi{10.3847/1538-4357/abf92c}

\bibitem[{{Fox}(2017)}]{psp2017}
{Fox}, N.~J. 2017, in AGU Fall Meeting Abstracts, Vol. 2017, SH21C--02

\bibitem[{{Garc\'{\i}a Marirrodriga, C.} {et~al.}(2021){Garc\'{\i}a
  Marirrodriga, C.}, {Pacros, A.}, {Strandmoe, S.}, {Arcioni, M.}, {Arts, A.},
  {Ashcroft, C.}, {Ayache, L.}, {Bonnefous, Y.}, {Brahimi, N.}, {Cipriani, F.},
  {Damasio, C.}, {De Jong, P.}, {D\'eprez, G.}, {Fahmy, S.}, {Fels, R.},
  {Fiebrich, J.}, {Hass, C.}, {Hern\'andez, C.}, {Icardi, L.}, {Junge, A.},
  {Kletzkine, P.}, {Laget, P.}, {Le Deuff, Y.}, {Liebold, F.}, {Lodiot, S.},
  {Marliani, F.}, {Mascarello, M.}, {M\"uller, D.}, {Oganessian, A.}, {Olivier,
  P.}, {Palombo, E.}, {Philippe, C.}, {Ragnit, U.}, {Ramachandran, J.},
  {S\'anchez P\'erez, J. M.}, {Stienstra, M. M.}, {Th\"urey, S.}, {Urwin, A.},
  {Wirth, K.}, \& {Zouganelis, I.}}]{Garc2021}
{Garc\'{\i}a Marirrodriga, C.}, {Pacros, A.}, {Strandmoe, S.}, {et~al.} 2021,
  A\&A, 646, A121, \dodoi{10.1051/0004-6361/202038519}

\bibitem[{Gary(2023)}]{Gary2023_review}
Gary, D.~E. 2023, Annual Review of Astronomy and Astrophysics, 61, null,
  \dodoi{10.1146/annurev-astro-071221-052744}

\bibitem[{{Gary} {et~al.}(2012){Gary}, {Nita}, \& {Sane}}]{Gary2012}
{Gary}, D.~E., {Nita}, G.~M., \& {Sane}, N. 2012, in American Astronomical
  Society Meeting Abstracts, Vol. 220, American Astronomical Society Meeting
  Abstracts \#220, 204.30

\bibitem[{{Hallinan} {et~al.}(2023){Hallinan}, {Anderson}, {Isella}, {Gary},
  {Bowman}, {Romero-Wolf}, \& {OVRO-LWA Collaboration}}]{Hallinan2023}
{Hallinan}, G., {Anderson}, M., {Isella}, A., {et~al.} 2023, in American
  Astronomical Society Meeting Abstracts, Vol.~55, American Astronomical
  Society Meeting Abstracts, 451.09

\bibitem[{{Hannah} \& {Kontar}(2012)}]{hannah2012}
{Hannah}, I.~G., \& {Kontar}, E.~P. 2012, \aap, 539, A146,
  \dodoi{10.1051/0004-6361/201117576}

\bibitem[{{Hannah} \& {Kontar}(2013)}]{hannah2013}
---. 2013, \aap, 553, A10, \dodoi{10.1051/0004-6361/201219727}

\bibitem[{Harris {et~al.}(2020)Harris, Millman, van~der Walt, Gommers,
  Virtanen, Cournapeau, Wieser, Taylor, Berg, Smith, Kern, Picus, Hoyer, van
  Kerkwijk, Brett, Haldane, del R{\'i}o, Wiebe, Peterson, G{\'e}rard-Marchant,
  Sheppard, Reddy, Weckesser, Abbasi, Gohlke, \& Oliphant}]{Harris2020}
Harris, C.~R., Millman, K.~J., van~der Walt, S.~J., {et~al.} 2020, Nature, 585,
  357, \dodoi{10.1038/s41586-020-2649-2}

\bibitem[{{Heywood} {et~al.}(2022){Heywood}, {Rammala}, {Camilo}, {Cotton},
  {Yusef-Zadeh}, {Abbott}, {Adam}, {Adams}, {Aldera}, {Asad}, {Bauermeister},
  {Bennett}, {Bester}, {Bode}, {Botha}, {Botha}, {Brederode}, {Buchner},
  {Burger}, {Cheetham}, {de Villiers}, {Dikgale-Mahlakoana}, {du Toit},
  {Esterhuyse}, {Fanaroff}, {February}, {Fourie}, {Frank}, {Gamatham}, {Geyer},
  {Goedhart}, {Gouws}, {Gumede}, {Hlakola}, {Hokwana}, {Hoosen}, {Horrell},
  {Hugo}, {Isaacson}, {J{\'o}zsa}, {Jonas}, {Joubert}, {Julie}, {Kapp},
  {Kenyon}, {Kotz{\'e}}, {Kriek}, {Kriel}, {Krishnan}, {Lehmensiek},
  {Liebenberg}, {Lord}, {Lunsky}, {Madisa}, {Magnus}, {Mahgoub}, {Makhaba},
  {Makhathini}, {Malan}, {Manley}, {Marais}, {Martens}, {Mauch}, {Merry},
  {Millenaar}, {Mnyandu}, {Mokone}, {Monama}, {Mphego}, {New}, {Ngcebetsha},
  {Ngoasheng}, {Ockards}, {Oozeer}, {Otto}, {Passmoor}, {Patel}, {Peens-Hough},
  {Perkins}, {Ramaila}, {Ramanujam}, {Ramudzuli}, {Ratcliffe}, {Robyntjies},
  {Salie}, {Sambu}, {Schollar}, {Schwardt}, {Schwartz}, {Serylak}, {Siebrits},
  {Sirothia}, {Slabber}, {Smirnov}, {Sofeya}, {Taljaard}, {Tasse}, {Tiplady},
  {Toruvanda}, {Twum}, {van Balla}, {van der Byl}, {van der Merwe}, {Van
  Tonder}, {Van Wyk}, {Venter}, {Venter}, {Wallace}, {Welz}, {Williams}, \&
  {Xaia}}]{heywood2022}
{Heywood}, I., {Rammala}, I., {Camilo}, F., {et~al.} 2022, The Astrophysical
  Journal, 925, 165, \dodoi{10.3847/1538-4357/ac449a}

\bibitem[{Hunter(2007)}]{Hunter:2007}
Hunter, J.~D. 2007, Computing in Science \& Engineering, 9, 90,
  \dodoi{10.1109/MCSE.2007.55}

\bibitem[{{Jonas} \& {MeerKAT Team}(2016)}]{meerkat2016}
{Jonas}, J., \& {MeerKAT Team}. 2016, in MeerKAT Science: On the Pathway to the
  SKA, 1, \dodoi{10.22323/1.277.0001}

\bibitem[{{Kassim} {et~al.}(2010){Kassim}, {White}, {Rodriquez}, {Hartman},
  {Hicks}, {Lazio}, {Stewart}, {Craig}, {Taylor}, {Cormier}, {Romero}, \&
  {Jenet}}]{kassim2010}
{Kassim}, N., {White}, S., {Rodriquez}, P., {et~al.} 2010, in Advanced Maui
  Optical and Space Surveillance Technologies Conference, ed. S.~{Ryan}, E59

\bibitem[{{Lemen} {et~al.}(2012){Lemen}, {Title}, {Akin}, {Boerner}, {Chou},
  {Drake}, {Duncan}, {Edwards}, {Friedlaender}, {Heyman}, {Hurlburt}, {Katz},
  {Kushner}, {Levay}, {Lindgren}, {Mathur}, {McFeaters}, {Mitchell}, {Rehse},
  {Schrijver}, {Springer}, {Stern}, {Tarbell}, {Wuelser}, {Wolfson}, {Yanari},
  {Bookbinder}, {Cheimets}, {Caldwell}, {Deluca}, {Gates}, {Golub}, {Park},
  {Podgorski}, {Bush}, {Scherrer}, {Gummin}, {Smith}, {Auker}, {Jerram},
  {Pool}, {Soufli}, {Windt}, {Beardsley}, {Clapp}, {Lang}, \&
  {Waltham}}]{Lemen2012}
{Lemen}, J.~R., {Title}, A.~M., {Akin}, D.~J., {et~al.} 2012, \solphys, 275,
  17, \dodoi{10.1007/s11207-011-9776-8}

\bibitem[{{Lonsdale} {et~al.}(2009){Lonsdale}, {Cappallo}, {Morales}, {Briggs},
  {Benkevitch}, {Bowman}, {Bunton}, {Burns}, {Corey}, {Desouza}, {Doeleman},
  {Derome}, {Deshpande}, {Gopala}, {Greenhill}, {Herne}, {Hewitt}, {Kamini},
  {Kasper}, {Kincaid}, {Kocz}, {Kowald}, {Kratzenberg}, {Kumar}, {Lynch},
  {Madhavi}, {Matejek}, {Mitchell}, {Morgan}, {Oberoi}, {Ord},
  {Pathikulangara}, {Prabu}, {Rogers}, {Roshi}, {Salah}, {Sault}, {Shankar},
  {Srivani}, {Stevens}, {Tingay}, {Vaccarella}, {Waterson}, {Wayth}, {Webster},
  {Whitney}, {Williams}, \& {Williams}}]{lonsdale2009}
{Lonsdale}, C.~J., {Cappallo}, R.~J., {Morales}, M.~F., {et~al.} 2009, IEEE
  Proceedings, 97, 1497, \dodoi{10.1109/JPROC.2009.2017564}

\bibitem[{{McMullin} {et~al.}(2007){McMullin}, {Waters}, {Schiebel}, {Young},
  \& {Golap}}]{mcmullin2007}
{McMullin}, J.~P., {Waters}, B., {Schiebel}, D., {Young}, W., \& {Golap}, K.
  2007, in Astronomical Society of the Pacific Conference Series, Vol. 376,
  Astronomical Data Analysis Software and Systems XVI, ed. R.~A. {Shaw},
  F.~{Hill}, \& D.~J. {Bell}, 127

\bibitem[{{Meadows}(2007)}]{Meadows2007}
{Meadows}, P. 2007, Journal of the British Astronomical Association, 117, 35

\bibitem[{{M\"uller, D.} {et~al.}(2020){M\"uller, D.}, {St. Cyr, O. C.},
  {Zouganelis, I.}, {Gilbert, H. R.}, {Marsden, R.}, {Nieves-Chinchilla, T.},
  {Antonucci, E.}, {Auch\`ere, F.}, {Berghmans, D.}, {Horbury, T. S.}, {Howard,
  R. A.}, {Krucker, S.}, {Maksimovic, M.}, {Owen, C. J.}, {Rochus, P.},
  {Rodriguez-Pacheco, J.}, {Romoli, M.}, {Solanki, S. K.}, {Bruno, R.},
  {Carlsson, M.}, {Fludra, A.}, {Harra, L.}, {Hassler, D. M.}, {Livi, S.},
  {Louarn, P.}, {Peter, H.}, {Sch\"uhle, U.}, {Teriaca, L.}, {del Toro Iniesta,
  J. C.}, {Wimmer-Schweingruber, R. F.}, {Marsch, E.}, {Velli, M.}, {De Groof,
  A.}, {Walsh, A.}, \& {Williams, D.}}]{muller2020}
{M\"uller, D.}, {St. Cyr, O. C.}, {Zouganelis, I.}, {et~al.} 2020, A\&A, 642,
  A1, \dodoi{10.1051/0004-6361/202038467}

\bibitem[{{Nakariakov} {et~al.}(2015){Nakariakov}, {Bisi}, {Browning}, {Maia},
  {Kontar}, {Oberoi}, {Gallagher}, {Cairns}, \&
  {Ratcliffe}}]{Nakariakov2015_SKA}
{Nakariakov}, V., {Bisi}, M.~M., {Browning}, P.~K., {et~al.} 2015, in Advancing
  Astrophysics with the Square Kilometre Array (AASKA14), 169,
  \dodoi{10.22323/1.215.0169}

\bibitem[{{Nindos}(2020)}]{Nindos2020}
{Nindos}, A. 2020, Frontiers in Astronomy and Space Sciences, 7, 57,
  \dodoi{10.3389/fspas.2020.00057}

\bibitem[{{Nindos} {et~al.}(2019){Nindos}, {Kontar}, \&
  {Oberoi}}]{Nindos2019_SKA}
{Nindos}, A., {Kontar}, E.~P., \& {Oberoi}, D. 2019, Advances in Space
  Research, 63, 1404, \dodoi{10.1016/j.asr.2018.10.023}

\bibitem[{{Nindos} {et~al.}(2021){Nindos}, {Patsourakos}, {Alissandrakis}, \&
  {Bastian}}]{nindos2021}
{Nindos}, A., {Patsourakos}, S., {Alissandrakis}, C.~E., \& {Bastian}, T.~S.
  2021, \aap, 652, A92, \dodoi{10.1051/0004-6361/202141241}

\bibitem[{{Oberoi} {et~al.}(2023){Oberoi}, {Bisoi}, {Sasikumar Raja},
  {Kansabanik}, {Mohan}, {Mondal}, \& {Sharma}}]{Oberoi2023}
{Oberoi}, D., {Bisoi}, S.~K., {Sasikumar Raja}, K., {et~al.} 2023, Journal of
  Astrophysics and Astronomy, 44, 40, \dodoi{10.1007/s12036-023-09917-z}

\bibitem[{{Offringa} {et~al.}(2014){Offringa}, {McKinley}, {Hurley-Walker},
  {Briggs}, {Wayth}, {Kaplan}, {Bell}, {Feng}, {Neben}, {Hughes}, {Rhee},
  {Murphy}, {Bhat}, {Bernardi}, {Bowman}, {Cappallo}, {Corey}, {Deshpande},
  {Emrich}, {Ewall-Wice}, {Gaensler}, {Goeke}, {Greenhill}, {Hazelton},
  {Hindson}, {Johnston-Hollitt}, {Jacobs}, {Kasper}, {Kratzenberg}, {Lenc},
  {Lonsdale}, {Lynch}, {McWhirter}, {Mitchell}, {Morales}, {Morgan},
  {Kudryavtseva}, {Oberoi}, {Ord}, {Pindor}, {Procopio}, {Prabu}, {Riding},
  {Roshi}, {Shankar}, {Srivani}, {Subrahmanyan}, {Tingay}, {Waterson},
  {Webster}, {Whitney}, {Williams}, \& {Williams}}]{Offringa2014}
{Offringa}, A.~R., {McKinley}, B., {Hurley-Walker}, N., {et~al.} 2014, Monthly
  Notices of the Royal Astronomical Society, 444, 606,
  \dodoi{10.1093/mnras/stu1368}

\bibitem[{Pick \& Vilmer(2008)}]{Pick2008}
Pick, M., \& Vilmer, N. 2008, The Astronomy and Astrophysics Review, 16, 1,
  \dodoi{10.1007/s00159-008-0013-x}

\bibitem[{Price-Whelan {et~al.}(2018)Price-Whelan, Sip{\H{o}}cz, G{\"u}nther,
  Lim, Crawford, Conseil, Shupe, Craig, Dencheva, Ginsburg,
  {et~al.}}]{price2018astropy}
Price-Whelan, A.~M., Sip{\H{o}}cz, B., G{\"u}nther, H., {et~al.} 2018, The
  Astronomical Journal, 156, 123

\bibitem[{Raouafi {et~al.}(2023)Raouafi, Matteini, Squire, Badman, Velli,
  Klein, Chen, Matthaeus, Szabo, Linton, Allen, Szalay, Bruno, Decker,
  Akhavan-Tafti, Agapitov, Bale, Bandyopadhyay, Battams, Ber{\v{c}}i{\v{c}},
  Bourouaine, Bowen, Cattell, Chandran, Chhiber, Cohen, D'Amicis, Giacalone,
  Hess, Howard, Horbury, Jagarlamudi, Joyce, Kasper, Kinnison, Laker, Liewer,
  Malaspina, Mann, McComas, Niembro-Hernandez, Nieves-Chinchilla, Panasenco,
  Pokorn{\'y}, Pusack, Pulupa, Perez, Riley, Rouillard, Shi, Stenborg,
  Tenerani, Verniero, Viall, Vourlidas, Wood, Woodham, \&
  Woolley}]{Raouafi2023}
Raouafi, N.~E., Matteini, L., Squire, J., {et~al.} 2023, Space Science Reviews,
  219, 8, \dodoi{10.1007/s11214-023-00952-4}

\bibitem[{Rast {et~al.}(2021)Rast, Bello~Gonz{\'a}lez, Bellot~Rubio, Cao,
  Cauzzi, DeLuca, De~Pontieu, Fletcher, Gibson, Judge, Katsukawa, Kazachenko,
  Khomenko, Landi, Mart{\'i}nez~Pillet, Petrie, Qiu, Rachmeler, Rempel,
  Schmidt, Scullion, Sun, Welsch, Andretta, Antolin, Ayres, Balasubramaniam,
  Ballai, Berger, Bradshaw, Campbell, Carlsson, Casini, Centeno, Cranmer,
  Criscuoli, DeForest, Deng, Erd{\'e}lyi, Fedun, Fischer,
  Gonz{\'a}lez~Manrique, Hahn, Harra, Henriques, Hurlburt, Jaeggli, Jafarzadeh,
  Jain, Jefferies, Keys, Kowalski, Kuckein, Kuhn, Kuridze, Liu, Liu, Longcope,
  Mathioudakis, McAteer, McIntosh, McKenzie, Miralles, Morton, Muglach, Nelson,
  Panesar, Parenti, Parnell, Poduval, Reardon, Reep, Schad, Schmit, Sharma,
  Socas-Navarro, Srivastava, Sterling, Suematsu, Tarr, Tiwari, Tritschler,
  Verth, Vourlidas, Wang, Wang, NSO, scientists, the DKIST Science
  Working~Group, \& the DKIST Critical Science Plan~Community}]{DKIST_2021}
Rast, M.~P., Bello~Gonz{\'a}lez, N., Bellot~Rubio, L., {et~al.} 2021, Solar
  Physics, 296, 70, \dodoi{10.1007/s11207-021-01789-2}

\bibitem[{{Reber}(1944)}]{reber1944}
{Reber}, G. 1944, \apj, 100, 279, \dodoi{10.1086/144668}

\bibitem[{{Rimmele} {et~al.}(2020){Rimmele}, {Warner}, {Keil}, {Goode},
  {Kn{\"o}lker}, {Kuhn}, {Rosner}, {McMullin}, {Casini}, {Lin}, {W{\"o}ger},
  {von der L{\"u}he}, {Tritschler}, {Davey}, {de Wijn}, {Elmore}, {Fehlmann},
  {Harrington}, {Jaeggli}, {Rast}, {Schad}, {Schmidt}, {Mathioudakis},
  {Mickey}, {Anan}, {Beck}, {Marshall}, {Jeffers}, {Oschmann}, {Beard},
  {Berst}, {Cowan}, {Craig}, {Cross}, {Cummings}, {Donnelly}, {de Vanssay},
  {Eigenbrot}, {Ferayorni}, {Foster}, {Galapon}, {Gedrites}, {Gonzales},
  {Goodrich}, {Gregory}, {Guzman}, {Guzzo}, {Hegwer}, {Hubbard}, {Hubbard},
  {Johansson}, {Johnson}, {Liang}, {Liang}, {McQuillen}, {Mayer}, {Newman},
  {Onodera}, {Phelps}, {Puentes}, {Richards}, {Rimmele}, {Sekulic}, {Shimko},
  {Simison}, {Smith}, {Starman}, {Sueoka}, {Summers}, {Szabo}, {Szabo},
  {Wampler}, {Williams}, \& {White}}]{dkist2020}
{Rimmele}, T.~R., {Warner}, M., {Keil}, S.~L., {et~al.} 2020, \solphys, 295,
  172, \dodoi{10.1007/s11207-020-01736-7}

\bibitem[{{Santander-Vela} {et~al.}(2021){Santander-Vela}, {Bartolini}, {and},
  {Miccolis}, \& {Rees}}]{SKAO2021}
{Santander-Vela}, J., {Bartolini}, M., {and}, {Miccolis}, M., \& {Rees}, N.
  2021, arXiv e-prints, arXiv:2110.13329, \dodoi{10.48550/arXiv.2110.13329}

\bibitem[{{The CASA Team} {et~al.}(2022){The CASA Team}, Bean, Bhatnagar,
  Castro, Meyer, Emonts, Garcia, Garwood, Golap, Villalba, Harris, Hayashi,
  Hoskins, Hsieh, Jagannathan, Kawasaki, Keimpema, Kettenis, Lopez, Marvil,
  Masters, McNichols, Mehringer, Miel, Moellenbrock, Montesino, Nakazato, Ott,
  Petry, Pokorny, Raba, Rau, Schiebel, Schweighart, Sekhar, Shimada, Small,
  Steeb, Sugimoto, Suoranta, Tsutsumi, van Bemmel, Verkouter, Wells, Xiong,
  Szomoru, Griffith, Glendenning, \& Kern}]{CASA2022}
{The CASA Team}, Bean, B., Bhatnagar, S., {et~al.} 2022, Publications of the
  Astronomical Society of the Pacific, 134, 114501,
  \dodoi{10.1088/1538-3873/ac9642}

\bibitem[{Tingay {et~al.}(2013)Tingay, Goeke, Bowman, Emrich, Ord, Mitchell,
  Morales, Booler, Crosse, Wayth, \& et~al.}]{tingay2013}
Tingay, S.~J., Goeke, R., Bowman, J.~D., {et~al.} 2013, Publications of the
  Astronomical Society of Australia, 30, e007, \dodoi{10.1017/pasa.2012.007}

\bibitem[{{van Haarlem} {et~al.}(2013){van Haarlem}, {Wise}, {Gunst}, {Heald},
  {McKean}, {Hessels}, {de Bruyn}, {Nijboer}, {Swinbank}, {Fallows},
  {Brentjens}, {Nelles}, {Beck}, {Falcke}, {Fender}, {H{\"o}randel},
  {Koopmans}, {Mann}, {Miley}, {R{\"o}ttgering}, {Stappers}, {Wijers},
  {Zaroubi}, {van den Akker}, {Alexov}, {Anderson}, {Anderson}, {van Ardenne},
  {Arts}, {Asgekar}, {Avruch}, {Batejat}, {B{\"a}hren}, {Bell}, {Bell}, {van
  Bemmel}, {Bennema}, {Bentum}, {Bernardi}, {Best}, {B{\^\i}rzan}, {Bonafede},
  {Boonstra}, {Braun}, {Bregman}, {Breitling}, {van de Brink}, {Broderick},
  {Broekema}, {Brouw}, {Br{\"u}ggen}, {Butcher}, {van Cappellen}, {Ciardi},
  {Coenen}, {Conway}, {Coolen}, {Corstanje}, {Damstra}, {Davies}, {Deller},
  {Dettmar}, {van Diepen}, {Dijkstra}, {Donker}, {Doorduin}, {Dromer}, {Drost},
  {van Duin}, {Eisl{\"o}ffel}, {van Enst}, {Ferrari}, {Frieswijk}, {Gankema},
  {Garrett}, {de Gasperin}, {Gerbers}, {de Geus}, {Grie{\ss}meier}, {Grit},
  {Gruppen}, {Hamaker}, {Hassall}, {Hoeft}, {Holties}, {Horneffer}, {van der
  Horst}, {van Houwelingen}, {Huijgen}, {Iacobelli}, {Intema}, {Jackson},
  {Jelic}, {de Jong}, {Juette}, {Kant}, {Karastergiou}, {Koers}, {Kollen},
  {Kondratiev}, {Kooistra}, {Koopman}, {Koster}, {Kuniyoshi}, {Kramer},
  {Kuper}, {Lambropoulos}, {Law}, {van Leeuwen}, {Lemaitre}, {Loose}, {Maat},
  {Macario}, {Markoff}, {Masters}, {McFadden}, {McKay-Bukowski}, {Meijering},
  {Meulman}, {Mevius}, {Middelberg}, {Millenaar}, {Miller-Jones}, {Mohan},
  {Mol}, {Morawietz}, {Morganti}, {Mulcahy}, {Mulder}, {Munk}, {Nieuwenhuis},
  {van Nieuwpoort}, {Noordam}, {Norden}, {Noutsos}, {Offringa}, {Olofsson},
  {Omar}, {Orr{\'u}}, {Overeem}, {Paas}, {Pandey-Pommier}, {Pandey}, {Pizzo},
  {Polatidis}, {Rafferty}, {Rawlings}, {Reich}, {de Reijer}, {Reitsma},
  {Renting}, {Riemers}, {Rol}, {Romein}, {Roosjen}, {Ruiter}, {Scaife}, {van
  der Schaaf}, {Scheers}, {Schellart}, {Schoenmakers}, {Schoonderbeek},
  {Serylak}, {Shulevski}, {Sluman}, {Smirnov}, {Sobey}, {Spreeuw}, {Steinmetz},
  {Sterks}, {Stiepel}, {Stuurwold}, {Tagger}, {Tang}, {Tasse}, {Thomas},
  {Thoudam}, {Toribio}, {van der Tol}, {Usov}, {van Veelen}, {van der Veen},
  {ter Veen}, {Verbiest}, {Vermeulen}, {Vermaas}, {Vocks}, {Vogt}, {de Vos},
  {van der Wal}, {van Weeren}, {Weggemans}, {Weltevrede}, {White}, {Wijnholds},
  {Wilhelmsson}, {Wucknitz}, {Yatawatta}, {Zarka}, {Zensus}, \& {van
  Zwieten}}]{lofar2013}
{van Haarlem}, M.~P., {Wise}, M.~W., {Gunst}, A.~W., {et~al.} 2013, Astronomy
  and Astrophysics, 556, A2, \dodoi{10.1051/0004-6361/201220873}

\bibitem[{{Wayth} {et~al.}(2018){Wayth}, {Tingay}, {Trott}, {Emrich},
  {Johnston-Hollitt}, {McKinley}, {Gaensler}, {Beardsley}, {Booler}, {Crosse},
  {Franzen}, {Horsley}, {Kaplan}, {Kenney}, {Morales}, {Pallot}, {Sleap},
  {Steele}, {Walker}, {Williams}, {Wu}, {Cairns}, {Filipovic}, {Johnston},
  {Murphy}, {Quinn}, {Staveley-Smith}, {Webster}, \& {Wyithe}}]{Wayth2018}
{Wayth}, R.~B., {Tingay}, S.~J., {Trott}, C.~M., {et~al.} 2018, Publications of
  the Astronomical Society of Australia, 35, e033, \dodoi{10.1017/pasa.2018.37}

\bibitem[{Zarka {et~al.}(2018)Zarka, Coffre, Denis, Dumez-Viou, Girard,
  Grießmeier, Loh, \& Tagger}]{Zarka2018}
Zarka, P., Coffre, A., Denis, L., {et~al.} 2018, in 2018 2nd URSI Atlantic
  Radio Science Meeting (AT-RASC), 1--1,
  \dodoi{10.23919/URSI-AT-RASC.2018.8471648}

\bibitem[{{Zirin} {et~al.}(1991){Zirin}, {Baumert}, \& {Hurford}}]{zirin1991}
{Zirin}, H., {Baumert}, B.~M., \& {Hurford}, G.~J. 1991, \apj, 370, 779,
  \dodoi{10.1086/169861}

\end{thebibliography}
\bibliographystyle{aasjournal}
\end{document}